\documentclass[12pt]{article}
\pdfoutput=1
\usepackage{graphics}
\usepackage{color}
\usepackage{listings}
\usepackage{fullpage}
\usepackage{graphicx}
\usepackage{rotating}
\usepackage{subfigure}
\usepackage{amsmath}
\usepackage{amssymb}
\usepackage{amsfonts}
\usepackage{mathrsfs}
\usepackage{latexsym}
\usepackage{authblk}
\usepackage{textpos}
\usepackage{bbm}
\usepackage{myoptions}
\usepackage{dsfont}
\usepackage{float}

\usepackage{bmpsize}

\numberwithin{equation}{section}

\pdfoutput=1

\begin{document}

\title{\bf Entanglement Growth after a Global Quench in Free Scalar Field Theory}

\author[1]{ Jordan S. Cotler}
\affil[1]{\em \small Stanford Institute for Theoretical Physics, 
Department of Physics, \protect\\
Stanford University, Stanford, CA 94305}

\author[2]{ Mark P. Hertzberg}
\affil[2]{\em \small Institute of Cosmology, Department of Physics and Astronomy, \protect\\
Tufts University, Medford, MA 02155}

\author[3]{\\  M\'ark Mezei}
\affil[3]{\em  \small Princeton Center for Theoretical Science, \protect\\
Princeton University, Princeton, NJ 08544}

\author[4]{ Mark T. Mueller}
\affil[4]{ \em \small Center for Theoretical Physics, Department of Physics, \protect\\
Massachusetts Institute of Technology, Cambridge, MA 02139}
\date{}

\maketitle

\begin{textblock*}{5cm}(11cm,-12.2cm)
\fbox{\footnotesize MIT-CTP-4827}
\end{textblock*}

\begin{abstract}
\noindent We compute the entanglement and R\'enyi entropy growth after a global quench in various dimensions in free scalar field theory. We study two types of quenches: a boundary state quench and a global mass quench.  Both of these quenches are investigated for a strip geometry in 1, 2, and 3 spatial dimensions, and for a spherical geometry in 2 and 3 spatial dimensions.  We compare the numerical results for massless free scalars in these geometries with the predictions of the analytical quasiparticle model based on EPR pairs, and find excellent agreement in the limit of large region sizes. At subleading order in the region size, we observe an anomalous  logarithmic growth of entanglement coming from the zero mode of the scalar. 

\let\thefootnote\relax\footnotetext{\hspace{-0.75cm}{\tt $^1$jcotler@stanford.edu \\ $^2$mark.hertzberg@tufts.edu \\ $^3$mezei@princeton.edu \\ $^4$mark.t.mueller@mac.com}}
\end{abstract}

\newpage
\tableofcontents
\newpage

\section{Introduction}

A global quench is a simple setting in which we can study thermalization in isolated quantum systems: at $t=0$ we start with an atypical translationally invariant, short range entangled initial state $\ket{\psi_0}$, and let the state evolve in time.\footnote{We use the word quench somewhat loosely; a more narrow definition describes a process in which the abrupt change of the Hamiltonian turns the ground state of the pre-quench Hamiltonian into an excited state of the post-quench Hamiltonian.}  In a generic quantum system, during the process of thermalization all simple observables converge to the value they take in the Gibbs ensemble.
A good characterization of thermalization is how close the reduced density matrix of a small subsystem $\sA$, $\rho_\sA\le[\ket{\psi(t)}\ri]$ is to the reduction of the thermal density matrix to the region $\sA$, $\rho_\sA^\text{th}\propto \Tr_{\bar \sA} e^{-\beta H}$, where $\beta$ is to be chosen such that the expectation value of the energy agrees between the two density matrices. One way to quantify the proximity to thermal behavior is to calculate the von Neumann entropy of $\rho_\sA(t)$, and follow as it evolves from an area law value to saturation at the thermal entropy. 

In a free theory, because of the infinitely many conserved charges the above picture requires modification. The time evolution leads to simple observables converging to their values in the  Generalized Gibbs Ensemble (GGE)~\cite{rigol2007relaxation} instead of the Gibbs ensemble. In this paper we will work with Gaussian states in free scalar field theories: for these states it is known that the set of charges that one has to include in the GGE are the particle numbers in each momentum mode~\cite{Sotiriadis:2014uza}.\footnote{For non-Gaussian states the story is more complicated~\cite{Cardy:2015xaa,Sotiriadis:2015xia}.} After the quench we focus on the case of massless fields.

We investigate entanglement entropy growth of these fields for geometric subregions in diverse dimensions. We discretize the theory on a lattice and use the correlator matrix approach to numerically compute entanglement and R\'enyi entropies. The continuum limit can be achieved by taking a scaling limit:
\es{Scaling1}{
	 {R\ov a}\,,{t\ov a} \gg 1\,, \qquad {\beta_k\ov a} \gg 1\,,
}
where $R$ is the characteristic size of the region $\sA$, $t$ is the time measured from the quench, $a$ is the lattice constant, and $\beta_k$ is the inverse of the effective temperature in the mode with wavenumber $k$. 
Let us introduce 
\es{Shat}{
	\hat{S}_\sA(t) = S_\sA(t) - S_\sA(0) 
}
to get rid of the vacuum area law pieces in the entropy.\footnote{We want the subtracted entropy $\hat{S}_\sA(t)$ to have a good continuum limit. In theories with low-dimension scalar operators the entropy can exhibit a state-dependent divergence structure~\cite{Marolf:2016dob}. In these theories there should exist a corresponding ambiguity in the definition of the entropy that allows us to regularize the entropy in a way that $\hat{S}_\sA(t)$ is finite for all times. In theories with state-independent divergence structure any regularization will yield a finite result. Free scalar theories are of the latter type. Of course, $S_\sA(t)$ itself is well-defined on the lattice. }
In the limit of large region sizes and times
\es{Scaling2}{
	 {R,t} \gg \beta_k\,, 
}
it is expected that the entropy obeys a scaling form:
\es{ScalingLaw}{
\hat{S}_\sA(t)&=s\,\vol(\sA)\, f\le(t\ov R\ri)\\
f(0)&=0\,, \qquad f(\infty)=1 \,,
}
where $s$ is the entropy density in the GGE,\footnote{In a generic system without any conserved quantity other than the energy, it would be the thermal entropy density.} $f(0)=0$ follows from the definition, and $f(\infty)=1$ assumes that the entropy reaches the equilibrium value predicted by GGE. In the limits~\eqref{Scaling1},~\eqref{Scaling2} the finite area law pieces in the entropy are suppressed by the factor $R/\beta$. In summary, we want to work in the double scaling limit
\es{Scaling3}{
	 {R,t} \gg \beta_k\gg a\,. 
}

\begin{figure}[!h]
\begin{center}
\includegraphics[width=5cm]{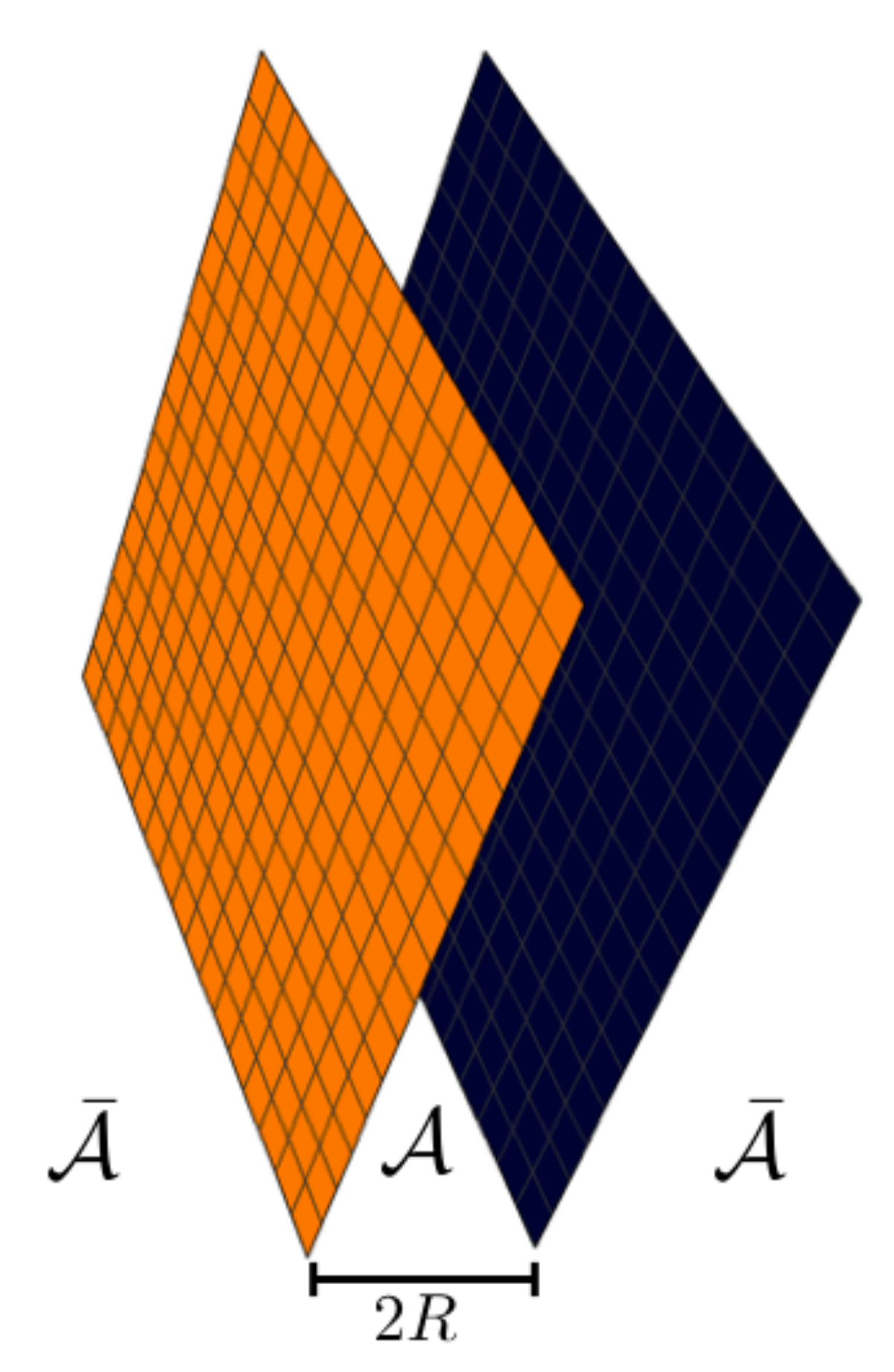}\,\,\,\,\,\,\,\,\,\,\,\,\,\,\,\,\,\,\,\,\,\,\,
\includegraphics[width=6cm]{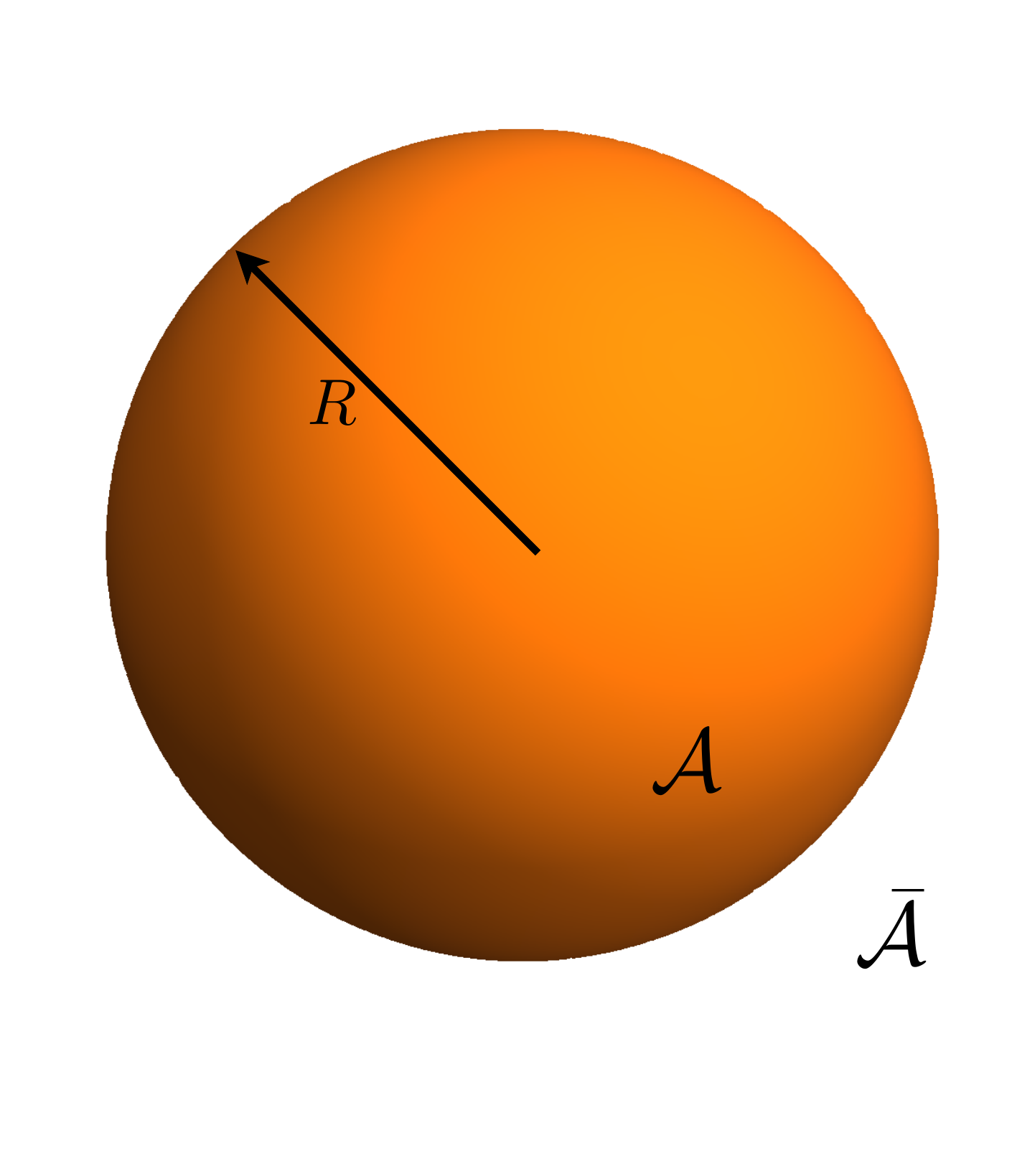}
\caption{\small{The two types of geometries we examine in this work. Regions $\mathcal{A}$ and $\bar{\mathcal{A}}$ partition the system into two distinct regions. Starting with a pure state, we trace out region $\bar{\mathcal{A}}$ to obtain a reduced density matrix $\rho_{\mathcal{A}}$, from which we compute the entanglement and R\'enyi entropies. Left: The strip geometry with two sides separated by a distance $2R$. Right: A spherical geometry of radius $R$.}
\label{fig:geometry}}
\end{center}
\end{figure}

There is a useful toy model for entanglement growth introduced in~\cite{Calabrese:2005in,Calabrese:2007rg} and generalized to higher dimensions in~\cite{Casini:2015zua}. This model assumes that the quench creates quasiparticle EPR pairs\footnote{In higher dimensions we can consider more complicated patterns of entanglement, as explored in~\cite{Casini:2015zua}. Intuitively, however, for Gaussian states that we consider in this paper the bipartite entanglement structure encoded in EPR pairs seems to be the appropriate choice.} localized on length scales $O(\beta)$. In the scaling limit~\eqref{Scaling3} the pairs can be taken to be pointlike. In a massless theory, the pairs then propagate with the speed of light,\footnote{In massive integrable models, they follow a nontrivial dispersion relation~\cite{fagotti2008evolution,2016arXiv160800614A}.}  and $\hat{S}_\sA(t)$ counts the number of pairs that have one member in $\sA$ and the other in $\bar{\sA}$. While the model reproduces the entropy of one interval in any $1+1$ dimensional conformal field theory (CFT)~\cite{Calabrese:2005in}, for more complicated geometries it only works in integrable CFTs~\cite{Asplund:2015eha}. In this work we find overwhelming evidence that the quasiparticle model reproduces the growth of entanglement in higher dimensional free massless scalar field theories in the scaling limit~\eqref{Scaling3}, by comparing the predictions of the quasiparticle picture to numerical computations in strip and sphere geometries, see Fig.~\ref{fig:geometry}. We study two types of quenches: the boundary state quench corresponds to starting the evolution from a regularized boundary state of the CFT, which leads to $\beta_k=\beta$, while the mass quench corresponds to abruptly changing the Hamiltonian of the system by changing the mass parameter, and leads to a $k$-dependent effective temperature. The quasiparticle picture works for both quenches equally well.

We emphasize three key features of our findings. First, we find (in the two geometries we consider) that at early times the entropy exhibits linear growth of the form:
\es{LinGrowth}{
	\hat{S}_\sA(t) &= v_E\,s\, \text{area}(\sA)\,t\,, \quad \beta_k\ll t\ll R \,,
}
where by $\text{area}(\sA)$ we mean the area of the entangling surface, $\vol(\p\sA)$. The dependence on the shape only appears through $\text{area}(\sA)$, and the entanglement velocity $v_E$ is shape independent.\footnote{In the regime $\beta_k\ll t\ll R$ the curvature of $\sA$ should be irrelevant for the process, so~\eqref{LinGrowth} is intuitive.~\eqref{LinGrowth} is also known to hold in strongly coupled theories with a holographic dual~\cite{Hartman:2013qma,Liu:2013iza,Liu:2013qca}.} Second, we comment on the saturation time $t_S$. For spherical geometries entanglement saturates as fast as allowed by causality~\cite{Hartman:2015apr},
\es{SatTime}{
	t_S^\text{(sphere)} = R \,.
}
We find that $t_S$ is strongly shape dependent,\footnote{In chaotic (holographic) examples the shape dependence of $t_S$ is mild, but still non-trivial~\cite{Liu:2013iza,Liu:2013qca,Mezei:2016wfz}.} and for a strip geometry\footnote{The intuition behind~\eqref{SatTime2} is that there are quasiparticle pairs propagating almost parallel to $\p\sA$ that take an arbitrary long time to start to contribute to the entropy. } 
\es{SatTime2}{
	t_S^\text{(strip)} = \infty \,.
}
We reiterate that the results~\eqref{LinGrowth},~\eqref{SatTime}, and~\eqref{SatTime2} are in agreement with the quasiparticle model. They are just simple properties of the function $\hat{S}_\sA(t)$ in the limit~\eqref{Scaling3}, which according to our findings is in complete agreement between the numerical computation in the free massless scalar field theory and the quasiparticle model.
Third, we point out an unexpected aspect of our numerical results: we see a logarithmic growth of entropy even after the saturation time~\eqref{SatTime}, which however is subleading in the limit~\eqref{Scaling3}, and therefore does not spoil the agreement with the quasiparticle model in the appropriate regime.\footnote{Unless we extrapolate this growth to exponentially large times.} We identify the scalar zero mode as the source of this logarithmic growth, but the phenomenon deserves further investigation.

Besides the intrinsic interest in the study of equilibration in free field theory, we are also motivated by the scarcity of computations of entropy growth in field theories. Our results elevate the status of the higher dimensional quasiparticle model from a toy model to an actual description of entanglement growth in free massless scalar field theories.\footnote{In $1+1$ dimensions the quasiparticle model has already been solidly established as a valid description of entanglement growth in integrable models.}\textsuperscript{,}\footnote{Of course, the outstanding problem is to give an analytic derivation of the quasiparticle picture from the field theory.} The results then provide a useful benchmark for strongly coupled theories: the conclusion of~\cite{Casini:2015zua} that in strongly interacting (holographic) theories entanglement spreads faster than allowed by free streaming,
\es{vEsmall}{
	v_E^\text{(free)} < v_E^\text{(holographic)}\,,
} 
is reinforced.  In general, collecting results on entanglement growth from various systems could lead to further insight into the workings of equilibration in quantum systems, both integrable and chaotic. For further discussion from this viewpoint see~\cite{Mezei:2016wfz}.

Using similar techniques, it is possible to study global quenches in free fermion theories.  The analytical and numerical techniques for analyzing global quenches in free scalar fields could potentially be extended to interacting field theories either perturbatively \cite{Hertzberg:2016eescalar} or non-perturbatively \cite{Cotler:2016ee1, Cotler:2016ee2}.  Such generalizations could shed new light on the dynamics of entanglement in interacting systems. In this paper we restrict our attention to instantenous quenches. It would be very interesting to extend our analysis to smooth quenches, where the duration of the quench $\de t$ introduces a new time scale. In the limit $R,t,\de t\gg \beta$, we expect the entropy to again obey a scaling form~\eqref{ScalingLaw}, but the scaling would become a function of two variables $f(t/R,\, \de t/R)$. Correlation functions obey universal scaling laws in this limit~\cite{Das:2014jna,Das:2014hqa,Berenstein:2014cia,Das:2015jka}, and it would be interesting to explore, if those results carry over to the case of entanglement entropy. It would also be interesting to see, if a modification of the quasiparticle model could reproduce the scaling function  $f(t/R,\, \de t/R)$. Perhaps, smearing the time of origin of the EPR pairs could be a useful starting point~\cite{Casini:2015zua}. 

The plan of the paper is as follows. In Sec.~\ref{sec:setup} we provide an introduction to our setup: we review the correlation matrix approach of computing entropies, and we discuss the quenches considered, along with the quasiparticle model.  Sec.~\ref{sec:numres} contains the numerical results for the entanglement and R\'enyi entropies, and a comparison with the quasiparticle model gives excellent agreement. A brief investigation into the logarithmically growing mode is also included. Some further details of the setup are relegated to the Appendices.

\section{Time evolution of entanglement}
\label{sec:setup}

We consider the time evolution of a Gaussian wave function in free scalar field theory through a quench. 
What enables us to do the computation is that the time evolution of an arbitrary Gaussian initial state remains Gaussian in a free theory. The computation simplifies in the global quench setup due to the preservation of translational and rotational symmetry: the kernel of the Gaussian remains diagonal for all times in momentum space. 
We can then apply the machinery developed for Gaussian states in free field theories~\cite{audenaert2002entanglement,botero2004spatial,peschel2009reduced,Coser:2014gsa,2016PhRvB..94b4306L} that we review below.

\subsection{Gaussian wave function in free scalar field theory}

Let us consider the Hamiltonian for a free massive scalar in $d+1$ spacetime dimensions
\es{ContHam}{
	H = \frac{1}{2}\int d^d x  \le[\pi^2+ (\nabla {\phi})^2 +m^2 \phi^2 \ri]\,,
}
where $\pi$ is the canonical momentum for $\phi$. The Hamiltonian~\eqref{ContHam} can be discretized and written in a general form
\es{HamK}{
	H = \frac{1}{2}\sum_{i=1}^N \pi_{i}^{2}+\frac{1}{2}\sum_{i,j=1}^N \phi _{i} \, K_{ij} \, \phi_{j} \,.
}
 We will consider Gaussian wave functions \footnote{For the purposes of computing entropies this is the most general Gaussian state. Linear terms in the argument of the exponential (leading to non-vanishing one point functions) can be transformed away using local unitaries leading to no change in the entropies. } 
\es{WaveFn}{
	\psi(t) = N(t) \exp\le[-\ha \sum_{i,j=1}^N \phi _{i} \, \Om_{ij}(t) \, \phi_{j}  \ri]\,,
}
where $N(t)$ includes the normalization of the wave function and an overall time dependent, but $\phi_i$-independent phase. If the wave function is of the form~\eqref{WaveFn} at one instant in time, it remains of the same form for all times when evolved by~\eqref{HamK}. The ground state of the system is obtained by setting $\Om=\sqrt{K}$. 

Let us diagonalize $K$ by making the orthogonal transformation $O$ on the fields $\phi$ and canonical momenta $\pi$
\be
	q = O \, \phi  \,, \quad\quad\quad  p = O^T \, \pi \,,
\ee
then $K$ takes the form
\be
	K=O^T K_D \, O\,, \label{KDeq}
\ee
where $K_D$ is diagonal.
$K$ is the discretized version of the operator $-\nabla^2+m^2$ and $O$ is the discrete Fourier transform. Then it is clear that to describe translationally and rotationally invariant quenches, it suffices to restrict to the case where $\Om(t)$ commutes with $K$.  In terms of the new $q$ variables the Gaussian~\eqref{WaveFn} is diagonal: 
\es{WaveFn2}{
	\psi(t) = N(t) \exp\le[-\ha \sum_{i=1}^N f_i(t) \, q _{i}^2  \ri]\,.
}
A simple example to keep in mind is a mass quench: we prepare the initial state through an abrupt change of the Hamiltonian $H\big\vert_{m^2}\to H\big\vert_{m^2=0}$, which is implemented by the change $K_0 \to K=K_0-m^2 \mathbf{1}$. All of our claims above are readily verified for this case.

As a warm-up problem for the time evolution, let us consider a quench in which we change the frequency of a harmonic oscillator abruptly at $t=0$, from frequency $\Lam$ before the quench to $\lam$ after the quench. With initial state given by the pre-quench ground state 
\be
	\psi_0(q) = \le(\Lam\ov \pi\ri)^{1/4} \exp\le[-{\Lam\ov 2}q^2\ri] \,,
\ee
the solution to the time-dependent Schr\"odinger equation reduces to a complex Riccati equation for the kernel $\Omega(t)$, which can easily be solved by standard methods to give the post-quench wave function\footnote{There are two easy checks of this formula: at $t=0$ it gives back the initial Gaussian, and for $\Lam=\lam$ we get the ground state wave function with trivial time dependence.} 
\es{EvolvedGaussian}{
	\psi(t,q)  = N(t) \exp\le[-{\lam\ov 2}\le({\lam+\Lam-(\lam-\Lam) e^{-2i\lam t}\ov \lam+\Lam+(\lam-\Lam) e^{-2i\lam t}}\ri) q^2\ri]\,.
}
For the correlation matrix approach, we need to calculate all two-point functions.\footnote{One point functions vanish by construction.}  A straightforward computation gives
\es{OneOscTwoPoint}{
	Q &\equiv \bra{\psi(t,q)} q^2\ket{\psi(t,q)} = {1\ov 4 \lam^2\Lam}\le[ \Lam^2+\lam^2 - (\Lam^2-\lam^2)\cos \le(2\lam\, t\ri)\ri] \\
	P &\equiv \bra{\psi(t,q)} p^2\ket{\psi(t,q)} = {1 \ov 4\Lam}\le[\Lam^2+\lam^2 + (\Lam^2-\lam^2)\cos \le(2\lam\, t\ri)\ri] \\  
	R &\equiv \bra{\psi(t,q)} \frac12 \{q,p\}\ket{\psi(t,q)} = {\Lam^2-\lam^2\ov 4 \Lam\lam}\sin \le(2\lam\, t\ri)\ .
}
Note that we do not lose any information by considering $ \frac12\{q,p\}$ instead of $qp$, because $ \frac12\{q,p\}=qp-{i\ov 2}$.  An important physical quantity is how much energy we inject into the system when we quench this harmonic oscillator:
\be
	\bra{\psi(t,q)} H_\lam\ket{\psi(t,q)}=\bra{\psi(t,q)}\le( {p^2\ov2}+{\lam^2 q^2\ov 2}\ri)\ket{\psi(t,q)}={\Lam^2+\lam^2\ov 4 \Lam} \,. \label{injEnergy} 
\ee

Based on the solution~\eqref{EvolvedGaussian} for a single harmonic oscillator, we immediately see that for a collection of harmonic oscillators of the discretized scalar field theory, the initial and post-quench wave functions are
\es{EvolvedGaussian2}{
	\Psi_0(q) &= \prod_{i = 1}^N \le( \Lam_i\ov \pi \ri)^{1/4} \exp \le[ -{\Lam_i\ov 2}q_i^2 \ri]  \\
	\Psi(t,q) &= \prod_{i = 1}^N \psi(t,q_i)\,,
}
where $\psi(t,q)$ is given in~\eqref{EvolvedGaussian}.  In~\eqref{EvolvedGaussian2} it is understood that one should make the replacement $q\to q_i$, $\lam \to \lam_i$, and $\Lam \to \Lam_i$, where $\Lam^2_i$ are the eigenvalues of $\Om(0)$ that characterize the initial state, and $\lam_i^2$ are the diagonal elements of $K_D$. Recall that as discussed around~\eqref{KDeq},  $\Om(0)$ and $K$ can be simultaneously diagonalized.    To be completely explicit, we write out the $\phi$-dependent part of the wave function as
\es{EvolvedGaussian3}{
	\Psi(t,\phi) &= N(t)  \exp\le[-\ha \sum_{ij=1}^N \phi _{i} \, \Om_{ij}(t) \phi_{j}  \ri]\\
	\Om_{ij}(t) &\equiv \sum_{k=1}^N  \lam_k \, {\lam_k + \Lam_k - (\lam_k-\Lam_k) e^{-2i\lam_k t} \ov \lam_k + \Lam_k + (\lam_k-\Lam_k) e^{-2i\lam_k t}}\, O_{ki} \, O_{kj} \,.
}

\subsection{The correlation matrix approach to quenches}

For this wave function it is now easy to determine the two-point functions of the canonical variables. The generalization of the single harmonic oscillator results for the two-point functions \eqref{OneOscTwoPoint} is
\es{}{
	Q_{ij} &\equiv \bra{\psi} \phi_i \phi_j \ket{\psi }=\sum_{k=1}^N O_{ki} \, O_{kj}\bra{\psi} q^2_k\ket{\psi }\\
	P_{ij} &\equiv \bra{\psi } \pi_i \pi_j\ket{\psi }=\sum_{k=1}^N O_{ki} \, O_{kj}\bra{\psi} p^2_k\ket{\psi }\\
	R_{ij} &\equiv \bra{\psi } \frac12 \{\phi_i ,\pi_j\}\ket{\psi }=\sum_{k=1}^N O_{ki} \, O_{kj}\bra{\psi}\frac12 \{ q_k,\, p_k\}\ket{\psi }\ .
}
Let us introduce a vector of canonical variables in region $\mathcal{A}$ (we trace over $\bar{\mathcal{A}}$)
\es{CanVar}{
	\chi_I = \begin{pmatrix}
	\phi_i \\ \pi_i
	\end{pmatrix} \,,
}
where $i,j=1,\dots,n$ are restricted to region $\mathcal{A}$ and  $I,J=1,\dots,2n$. The canonical commutation relations read:
\es{CanComm}{
	\le[\chi_I,\, \chi_J\ri]&=i \, J_{IJ}\,, \qquad 
	J_{IJ}=\begin{pmatrix}
	0 & \mathbf{1}\\
	-\mathbf{1}& 0
\end{pmatrix}\,,
}
and the correlators can be collected into a $2n\times2n$ matrix~\cite{audenaert2002entanglement,botero2004spatial,Coser:2014gsa}:
\es{CorrelationMtx}{
	\Gamma_{IJ}&=\frac12\bra{\psi }\{\chi_I\, , \chi_J\}\ket{\psi }
	=\begin{pmatrix}
	Q_{ij} & R_{ij}\\
	R_{ji} & P_{ij}
	\end{pmatrix} \,.
}
Note that $\Gamma_{IJ}$ is a real symmetric positive definite matrix.  Such matrices can be brought to Williamson normal form, i.e.~there exists a symplectic matrix $M$ that diagonalizes them. $M$ implements a canonical transformation (that preserves~\eqref{CanComm}):
\es{CanonicalTf}{
    \tilde{\chi}&=M\chi 
    \,,\qquad MJM^T = J \,,\\
    \tilde\Gamma &= M \, \Gamma M^T = 
    \begin{pmatrix}
    	{\rm diag}\le(\ga_k\ri)& 0\\
    	0 & {\rm diag}\le(\ga_k\ri)
    \end{pmatrix}\,.
}
The easiest way of determining $\ga_k$ is to obtain the eigenvalues of the matrix $iJ\Ga$, which are $\{\pm \ga_k\}$.
We have now successfully mapped the problem to computing the entropy of $n$ harmonic oscillators at finite temperatures:
\es{betas}{
	\beta_k=\log{\ga_k+1/2\ov\ga_k-1/2}\,.
}
Thus the entropy $S=-\mbox{Tr}[\rho_{\mathcal{A}}\log\rho_{\mathcal{A}}]$ is
\be
	S = \sum_{k=1}^n \tilde S(\ga_k) \,, \,\,\,\,\,
	\tilde S(\ga) \equiv \le(\ga+\frac12\ri) \, \log \le(\ga+\frac12\ri) - \le(\ga-\frac12\ri) \, \log \le(\ga-\frac12\ri) \,,
\ee
and the R\'enyi entropies $S_q=-{1\ov q-1}\log\mbox{Tr}[\rho_{\mathcal{A}}^q]$ are
\be
	S_q = \sum_{k=1}^n \tilde S_q(\ga_k) \,,
	\qquad 
	\tilde S_q(\ga) \equiv {1\ov q - 1} \log\le[\le(\ga+\frac12\ri)^q - \le(\ga-\frac12\ri)^q \ri] \,.
\ee
In the symmetric geometries $\sA$ that we consider in this paper, the matrix $\Gamma$ is block diagonal. In the case of the strip, the different blocks are labelled by the momenta parallel to the entangling surface; in the case of a sphere, the labels are the angular momentum quantum numbers.  The matrix $\Gamma$ is block diagonal as two point functions do not mix different linear or angular momenta.  In these cases, the above steps can be performed block by block, and the entropy is just the sum of the contribution of each block. Details of the different coordinate systems can be found in the Appendices.

\subsection{Different types of quenches}\label{sec:Quenches}

As discussed in the introduction, in free theories there are infinitely many conserved charges, and equilibration only happens in the sense of the GGE. What this means for our purposes is that any mode can be quenched independently, and they have their own effective temperature.
We shall focus on the case in which after the quench the mass of the scalar field is zero (so $\lam(k) = \wk$), so in the continuum limit~\eqref{Scaling1} the time evolution is governed by a CFT.

\subsubsection{Boundary state quench}

From the point of view of effective temperatures mode by mode, a particularly nice state to consider
is the conformal boundary state model of a quench~\cite{Calabrese:2005in}
\es{BdyState}{
	\ket{\Psi_0}&=\exp\le[-{\beta\ov 4}\, H\ri]\, \ket{\text{Dirichlet}}\,,
}
where $\ket{\text{Dirichlet}}$ is the Dirichlet boundary state.  This state will have finite energy in any dimensions, with mode-independent inverse temperature $\beta_\wk=\beta$, and is specified by the relation
\es{BdyQuench}{
	\Lam(\wk)_\text{bdy}={\wk\ov \tanh\!\le({\beta\, \wk\ov 4}\ri)}\,.
}
The requirement that the quench is described by the continuum field theory translates into ${\beta/ a} \gg 1$, as discussed around~\eqref{Scaling3}.\footnote{  Rewriting the inequality as $\frac{1}{a} \gg \frac{1}{\beta}$, we intuitively want the energy scale of a thermal excitation, which is approximately $1/\beta$, to be far less than the highest energy excitations which can be supported by our lattice theory, which go as $1/a$.}

The R\'enyi entropy density arising from this quench is (in $d$ spatial dimensions)
\es{RenyiEntropyDensityBoundaryStateQuench}{
	s_q &= {1 \ov q - 1} \int {d^dk \ov (2\pi)^{d}}\ \le[\log\le(1 - e^{- \beta k q} \ri) - q \log\le(1-e^{-{\beta} k}\ri)\ri]  \\
	&= {(q^{d+1} - 1) \ov q^d (q - 1)}{ \zeta(d+1) \, \Ga \le(d+1\ov 2\ri) \ov \pi^{(d+1)/2}}\,{1\ov \beta^d}
}
leading to the entropy density
\es{EntropyDensityBoundaryStateQuench}{
	s = \lim_{q \rightarrow 1} s_q = {(d + 1) \, \zeta(d+1) \, \Ga\le(d+1 \ov 2\ri)\ov \pi^{(d+1)/2}} \, {1 \ov \beta^d}
	= \begin{cases}
		\frac{\pi}{3 \beta} \qquad & (d=1)  \\[4pt]
		\frac{3 \, \zeta(3)}{2 \pi \beta^2} \qquad & (d=2)  \\[4pt]
		\frac{2 \, \pi^2}{45 \beta^3} \qquad & (d=3)
	\end{cases} \,.
}

\subsubsection{Mass quench}

In contrast to the boundary state quench, the mass quench -- although it may seem more physical -- has less favorable properties. In particular, the relation
\es{MassQuench}{
	\Lam(\wk)_\text{mass}=\sqrt{\wk^2+m^2}\,.
}
produces a mode-dependent temperature~\cite{Sotiriadis:2010si}
\es{betaEff}{
	\beta_\wk&={4\ov\wk}{\rm arctanh}\le(\wk\ov \Lam(k)\ri)
	= \begin{cases}
		{4\ov m}\qquad &\wk\ll m  \\[4pt]
		{4\log\le(\wk/m\ri)\ov \wk} \qquad &\wk\gg m 
	\end{cases} \, ,
}
i.e., high energy modes have diverging effective temperature.\footnote{The average energy in a high energy mode is still low.} 
Now the inequalities in~\eqref{Scaling3} will not be satisfied for all $k$. Nevertheless, we can intuit that the weaker condition    
\es{mReq}{
	m\, a\ll 1
}
should guarantee that we stay close to the continuum limit, which corresponds to low energies. 
However, additional complications arise for $d \geq 3$, where the mass quench does not produce a finite energy state, as the total injected energy (in excess of the vacuum energy) is
\be
	\Delta E=\bra{\Psi(t,y)} (H-E_\text{vac})\ket{\Psi(t,y)} 
		= \sum_k {(\Lam_k - \lam_k)^2 \ov 4 \Lam_k} \,,
\ee
where we used~\eqref{injEnergy}.
In the continuum limit the change in energy density is
\es{CriticalDim}{
	\frac{\Delta E}{V} = \int {d^dk \ov (2\pi)^d} \ {(\Lam(k) - k)^2 \ov 4 \Lam(k)}
	\sim m^{d+1} \, \le( {\UV \ov m} \ri)^{d-3} + \dots \,,
}
where $\UV \gg m$ is some UV cutoff scale.  For spatial dimensions $d = 1, 2$ the mass quench produces a state with finite energy density as $\UV \rightarrow \infty$, while for $d \ge 3$ we encounter ultraviolet divergences; in particular in $d = 3$ we find a logarithmic divergence. We may summarize these results as
\es{Modified}{
	\frac{\Delta E}{V} &=\begin{cases}
	{m^2\ov 2\pi} \quad &(d = 1)  \\[4pt]
	{m^3\ov 6\pi} \quad &(d = 2)  \\[4pt]
	\ \infty \quad &(d \geq 3)
	\end{cases} \,.
}
The R\'enyi entropy density for the mass quench is given by \eqref{RenyiEntropyDensityBoundaryStateQuench} with $\beta\to\beta_k$, leading to
\begin{align}
\label{RenyiEntropyDensityMassQuench}
	s_q = 
	 \begin{cases}
	\frac{-q + \cot(\pi/4q)}{2(q-1)} \, m \quad & (d = 1)  \\[4pt]
	\frac{2 \gamma_E + \psi(1-1/2q) + \psi(1+1/2q) + 2 q (\log 4 - 1)}{16 \pi (q-1)}\, m^2 \quad & (d = 2)  \\[4pt]
	\frac{4q - 3 \cot(\pi/4q) + \cot(3\pi/4q)}{48\pi (q-1)}\, m^3 \quad & (d = 3)
	\end{cases} \,,
\end{align}
where $\gamma_E$ is the Euler-Mascheroni constant and $\psi(z)$ is the digamma function.  The above equations yield the entropy density
\begin{align}
\label{EntropyDensityMassQuench}
	s = \lim_{q\to 1} s_q  = 
		 \begin{cases}
	\frac14 (\pi-2) \, m \quad  (d = 1)  \\[4pt]
	{\log 2\ov 4\pi}\, m^2 \quad  \,\,\,\,\,\,\,\,\,\,(d = 2)  \\[4pt]
	{1\ov 12\pi}\, m^3 \quad \,\,\,\,\,\,\,\,\,\,\,\, (d = 3)
	\end{cases} \,,
\end{align}
and $s = \infty$ is divergent for spatial dimensions $d \geq 4$.  Note that in $d = 3$ the entropy density is finite, even though the energy density is infinite after a mass quench. 

Furthermore, one also anticipates a divergent area law contribution to the entropy in any number of dimensions. The entropy difference from before to after the quench~\eqref{Shat} is expected to result in an infinite area law correction for $d\geq3$. For example, for $d=3$ there should be a log divergence in the change of the area law contribution \cite{Hertzberg:2010uv}
\begin{equation}
	\Delta S_\text{area} = {A\,m^2\over24\pi}\log\!\left(\UV\over m\right).
\end{equation}
So we will only focus on mass quenches in 1 and 2 spatial dimensions.

\subsubsection{Other quenches}

The above formalism extends to any quench in which the kernel of $\Psi_0$ is diagonal in Fourier space, meaning it takes the form~\eqref{EvolvedGaussian2}. We have discussed boundary state and mass quenches, since they are perhaps the most physical examples, but an arbitrary choice of $\Lam(k)$ is allowed. If we want to preserve translation and rotation invariance, we only need to choose a function $\Lam=\Lam(|{\bf k}|)$. This can be parametrized by a type of mode dependent initial mass that we quench
\es{Modified2}{
	\Lam(\wk) &= \sqrt{\wk^2+m^2(\wk)} \,, 
}
for the initial wave function.\footnote{The initial wave function can be thought of as the ground state of a (non-local) Hamiltonian with dispersion relation~\eqref{Modified2}.}  Choosing a mass function that decays to zero $(m^2(\wk)\to 0)$ fast enough for large $\wk$ can make $\Delta E$ finite in any dimension.

Another generalization that we may consider is to replace the instantaneous quench with a smooth quench of duration $\de t$, which would render the energy density injected into the system by a mass quench finite~\cite{Das:2015jka}.

\subsection{The quasiparticle model for a quench}

In this section, we will review the dynamics of entanglement using the quasiparticle picture in which entanglement is carried by a uniform density of noninteracting EPR pairs~\cite{Calabrese:2005in,Calabrese:2007rg,Casini:2015zua}.  We assume that the two quasiparticles which comprise a pair travel in opposite directions at the speed of light, with an isotropic angular distribution.   

In the quasiparticle model one can first fix a point $\mathbf{x}$ and time $t$, and determine the contribution to the entropy of region $\sA$ coming from EPR pairs that originated from that point. These pairs will be positioned on a sphere of radius $t$ with center $\mathbf{x}$. Let us denote the part of this sphere incident in region $\sA$ by $L_{\sA}(\mathbf{x},t)$, and by $\mu[L_{\sA}(\mathbf{x},t)]$ the contribution to the entropy of this region. We have to sum over the points of origin to obtain the entropy of region $\sA$:
\es{AddEntropy}{
	S_{\sA}(t) = \int d^{d} x \, \mu[L_{\sA}(\mathbf{x},t)]\,.
}

For any region $\sB$ on a sphere, let us denote the set of antipodal points as $\sB'$, and the complement of this set on the sphere by $\overline{\sB'}$. Then $\mu[\sB]$ is given by
\es{LCCont}{
	\mu[\sB]=s\, {\vol\le(\sB\cap \overline{\sB'}\ri)\ov \vol(\text{sphere})}\,.
}
This formula is intuitive. If $\sB$ is contained within a hemisphere, then $\sB\cap \overline{\sB'}=\sB$, and the quasiparticles in $\sB$ will have pairs in $\sB'$, which lies entirely in $\bar{\sA}$. Then the formula counts the fraction of the EPR pairs ${\vol\le(\sB\ri)/ \vol(\text{sphere})}$, which have one member inside $\sA$ and the other outside.  If, however, $\sB$ is a bigger (or more complicated) region, taking $\vol\le(\sB\cap \overline{\sB'}\ri)$ instead of $\vol\le(\sB\ri)$ eliminates those pairs, whose both members are in $\sA$.

Although not discussed in~\cite{Casini:2015zua}, the computation presented here applies both to the entanglement and R\'enyi entropies with the appropriate entropy density $s$ (or $s_q$) used in~\eqref{LCCont}.

In~\cite{Casini:2015zua} the integral~\eqref{AddEntropy} was evaluated for symmetric entangling regions. Here we will need the results for $S_\sA(t)$ in $d=1,2,3$ spatial dimensions for a strip of width $L = 2R$ and a sphere of radius $R$.  In $d=1$, both cases degenerate to $\mathcal{A}$ being a finite interval of length $2R$.  For the the strip geometry in spatial dimension $d$, we have
\es{qpPrediction}{
	{S^{(d=1)}_\text{strip}(t)\ov s} &= 
	\begin{cases}
        		2 t \qquad & \qquad\qquad\qquad\qquad\qquad\qquad\quad (t < R)  \\[4pt]
        		2 R \qquad & \qquad\qquad\qquad\qquad\qquad\qquad\quad (t > R)
        \end{cases}  \\[4pt]
	{S^{(d=2)}_\text{strip}(t)\ov s \, L} &=
	\begin{cases}
		{4 \ov \pi} \, t \qquad & (t < R)  \\[4pt]
		{4 \ov \pi} \, \le[ t - \sqrt{t^2 - R^2} + R \arccos \le( R \ov t \ri) \ri] \qquad & (t > R)
	\end{cases}  \\[4pt]
	{S^{(d=3)}_\text{strip}(t)\ov s \, L^2} &=
	\begin{cases}
		t \qquad\qquad\qquad\qquad\qquad\qquad\qquad\qquad & (t < R)  \\[4pt]
		2R - {R^2 \ov t} \qquad\qquad\qquad\qquad\qquad\qquad & ( t > R)
	\end{cases} \,.
}
Likewise for spherical geometries in spatial dimension $d$,
\es{qpPredictionSphere}{
        {S^{(d=2)}_\text{sphere}(t) \ov s}
        &=\begin{cases}
       		2\le[t\sqrt{R^2-t^2}+R^2\arcsin\le(t\ov R\ri)\ri] \quad \ & (t < R)  \\[4pt]
        		\pi R^2 \quad \ & (t > R)
        \end{cases}  \\[4pt]
        {S^{(d=3)}_\text{sphere}(t) \ov s}
        &=\begin{cases}
        		2\pi \le[ R^2 \, t - {1 \ov 3} t^3 \ri] \qquad\qquad\qquad\quad\quad & (t < R)  \\[4pt]
        		{4\pi \ov 3} \, R^3\qquad\qquad\qquad\qquad\quad\quad & (t > R)
        \end{cases} \,.
}
The saturation times~\eqref{SatTime} and~\eqref{SatTime2} can be easily read off from these expressions. The expression for the entanglement velocity  in $d$ spatial dimensions is 
\es{vEQP}{
	v_E = {\Ga\le(d \ov 2\ri) \ov \sqrt{\pi} \, \Ga\le(d+ 1 \ov 2 \ri)} \,.
}
In the next section, we will  confirm the predictions of $S_\sA(t)$ (and hence for $t_S$ and $v_E$) with numerical simulations of global quenches of free scalar fields.

\section{Numerical results for strips and spheres}\label{sec:numres}

\subsection{Intervals in 1 spatial dimension}

\begin{figure}[H]
\begin{center}
\includegraphics[scale=0.4]{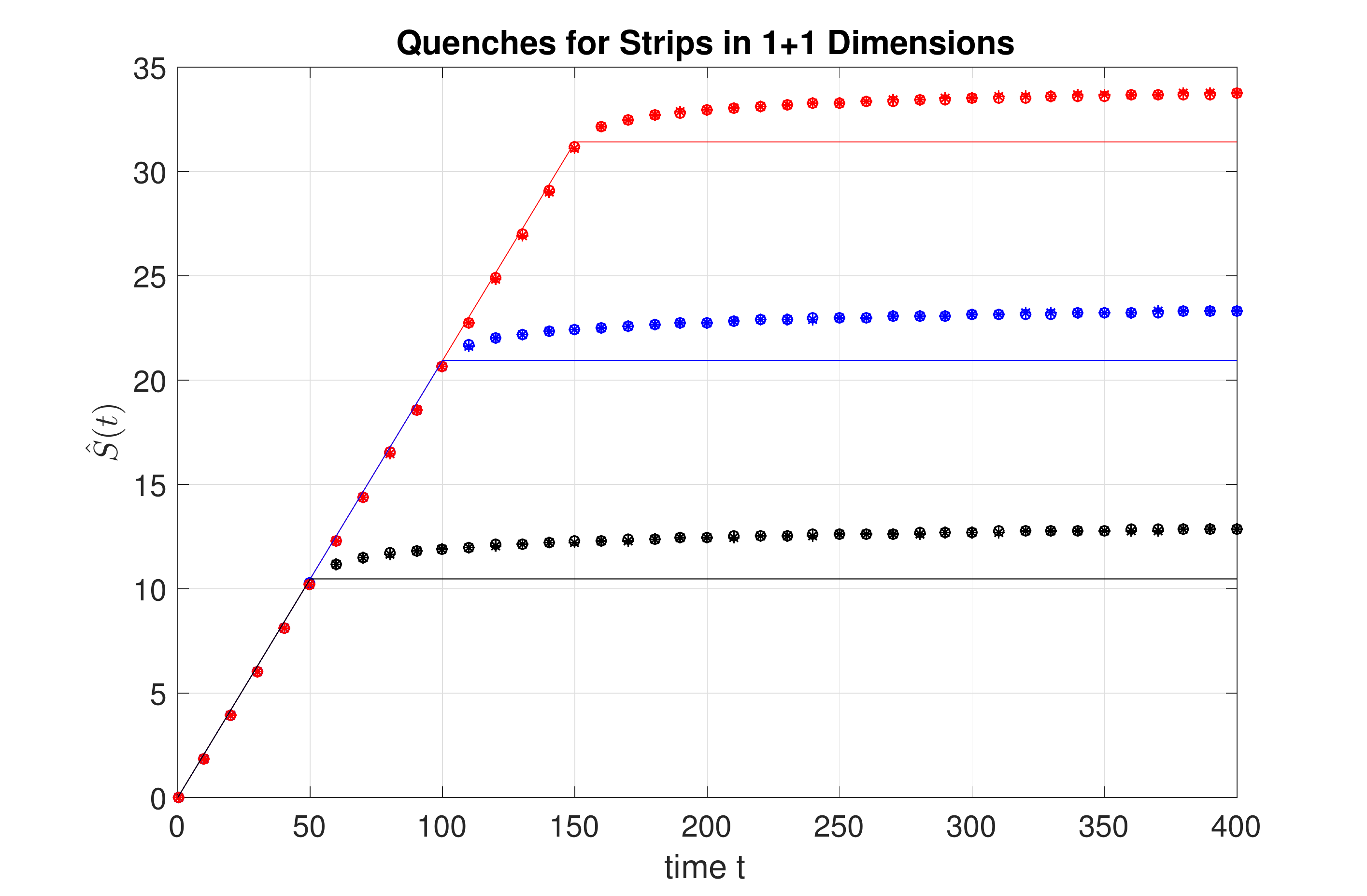}
\includegraphics[scale=0.4]{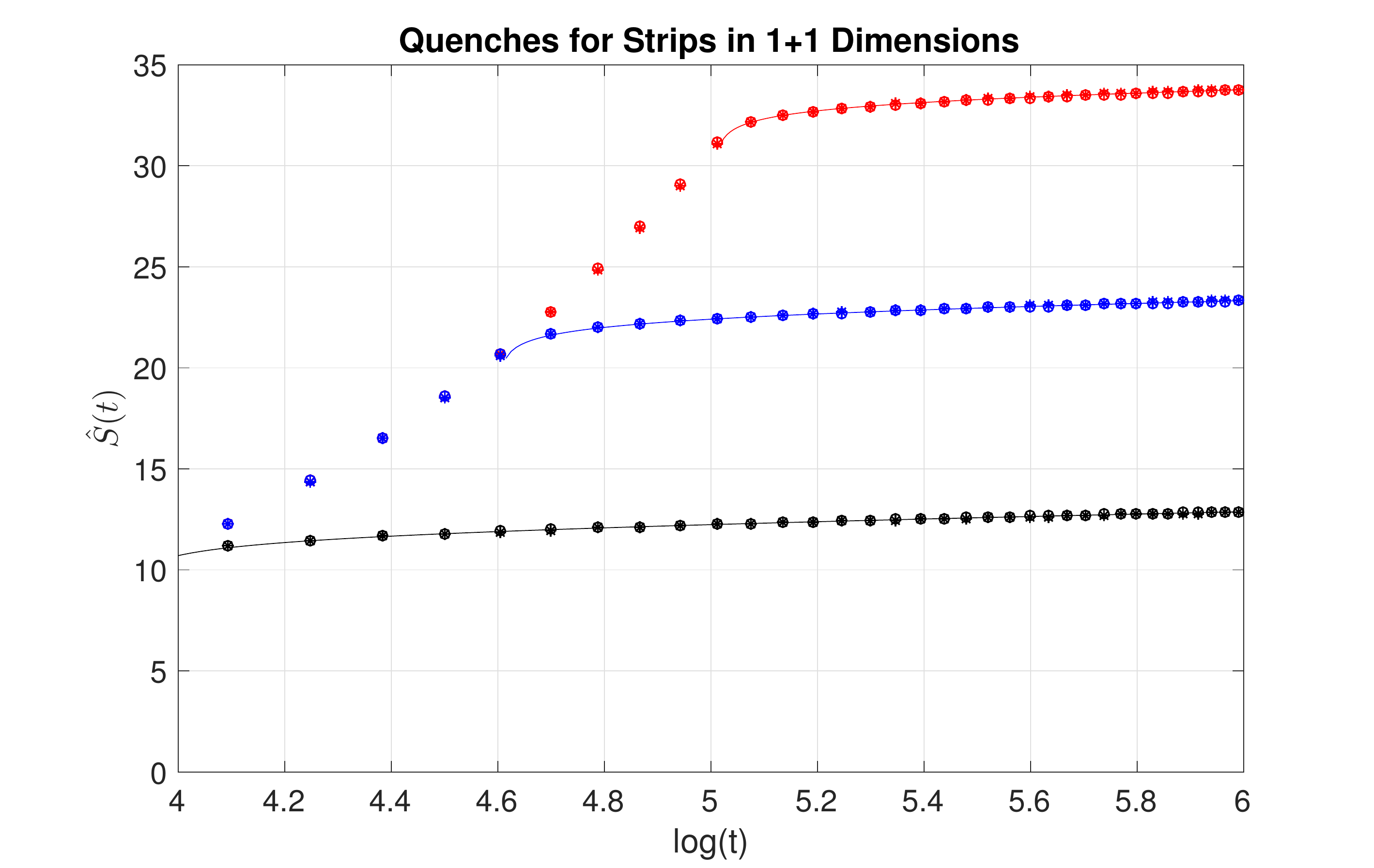}
\caption{\small{Quenches for strips (intervals) in 1+1 dimensions.  The top figure is linear time and the bottom figure is logarithmic time.  We have taken intervals of length $2R/a = 100, 200, 300$, where $a$ is the lattice unit, both for boundary state quenches with $\beta = 10 \, a$ and for mass quenches.  For the mass quench we have chosen $m ={4 \pi \ov 3(\pi-2) \beta}$ such that the resulting entropy density matches that of the boundary state quench, see~\eqref{RenyiEntropyDensityBoundaryStateQuench},~\eqref{EntropyDensityBoundaryStateQuench}.  Curves from lower to upper represent $2R/a = 100$ (black), $2R/a = 200$ (blue), and $2R/a = 300$ (red).  Curves with circle-markers represent boundary state quenches; curves with star-markers represent mass quenches (almost indistinguishable from the boundary state quenches).  In the top figure, the lines with no markers represent the quasiparticle model predictions~\eqref{qpPrediction}, using the entropy density corresponding to the quench.  Note that the linear ramp of the quasiparticle model is indistinguishable from the numerical results.  The lines in the bottom figure with no markers show the fit~\eqref{FitCurve} to the linear asymptotic behavior $\sim \frac12 \log(t)$.}
\label{fig:quenches_1d_strip}}
\end{center}
\end{figure}

In $d = 1$  spatial dimension  the results for the entropy for intervals of different sizes in a boundary state and a mass  quench  can be found in Fig.~\ref{fig:quenches_1d_strip}. For convenience we impose periodic boundary conditions at the ends of the 1 dimensional region.  In this figure we have used the subtracted entropy~\eqref{Shat}. Mass quenches for the similar systems were analyzed in~\cite{2014PhRvB..90t5438N,Coser:2014gsa}.  Related analytical and numerical results for local and global quenches in~\cite{peschel2009reduced} and references therein. 

In Fig.~\ref{fig:quenches_1d_strip} we see that the two types of quenches, the boundary quench and the mass quench, give results that are nearly indistinguishable to the eye. The quasiparticle prediction~\cite{Calabrese:2005in,Calabrese:2007rg} matches closely for $t  <  R$ with the correct entropy density $s$~\eqref{EntropyDensityBoundaryStateQuench}.  However, instead of sharp saturation at $t = R$, we see that the entropy keeps increasing as a logarithm of time.  To understand this deviation, we reproduce the same quenches in Fig.~\ref{fig:quenches_1d_strip} (bottom), but plotted in logarithmic time.  From fitting we find that the coefficient of the logarithm is independent of $R$, and appears to equal $1/2$.\footnote{In more detail, we have fitted
\es{FitCurve}{
	\hat{S}(t)=\sS+c\, \log\le(t-R\ov a\ri)
}
for the data points with $t > R$. \eqref{FitCurve} diverges for $t=R$, so we started the fitting procedure a couple of lattice units later in time. One can also consider introducing a time shift as an additional fitting parameter, but this hardly changes the value of $\sS$ and $c$. We found that $c=1/2$ within $2\%$ accuracy. On Fig.~\ref{fig:quenches_1d_strip} (bottom) we have plotted~\eqref{FitCurve} with $c=1/2$ and $\sS$ fitted. The match is excellent. }
  Since the coefficient of the logarithm is independent of $R$, in the limit~\eqref{Scaling3} this $\frac{1}{2}\log t$ behavior is subleading. Hence the prediction of the quasiparticle model is obeyed in the limit of large region size and time.\footnote{Unless we extrapolate this growth to exponentially large times.}  
  
\begin{figure}[H]
\begin{center}
\includegraphics[scale=0.4]{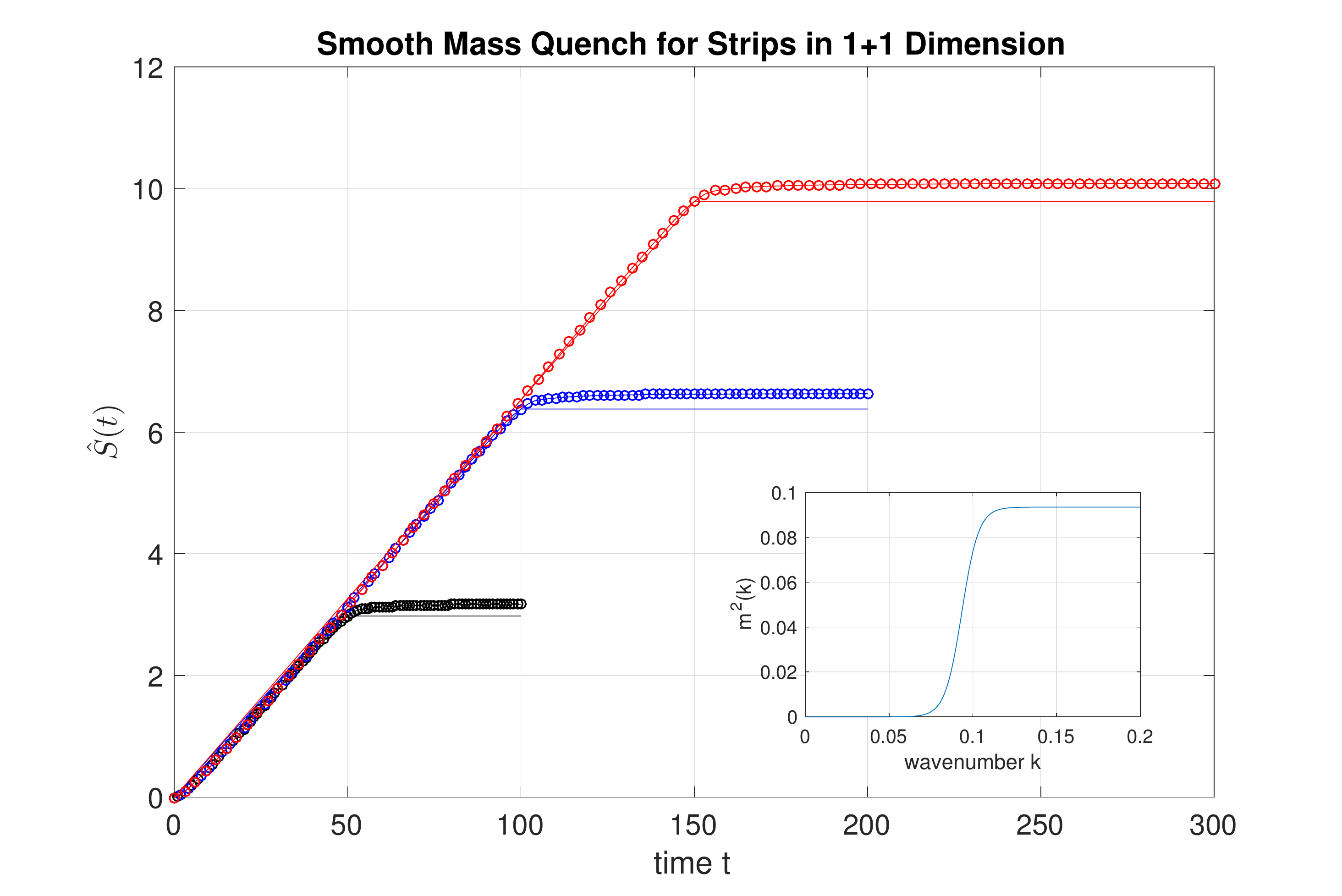}
\caption{\small{Modified mass quench for intervals of length $2R/a = 100, 200, 300$, with $m^2(k)$ a smoothed step function as shown in the inset graph.  $s$ in the quasiparticle formula~\eqref{qpPrediction} is adjusted to match the numerical data points.  
}}
\label{fig:strip_1d_sigmoid_quench}
\end{center}
\end{figure}

\begin{figure}[H]
\begin{center}
\includegraphics[scale=0.4]{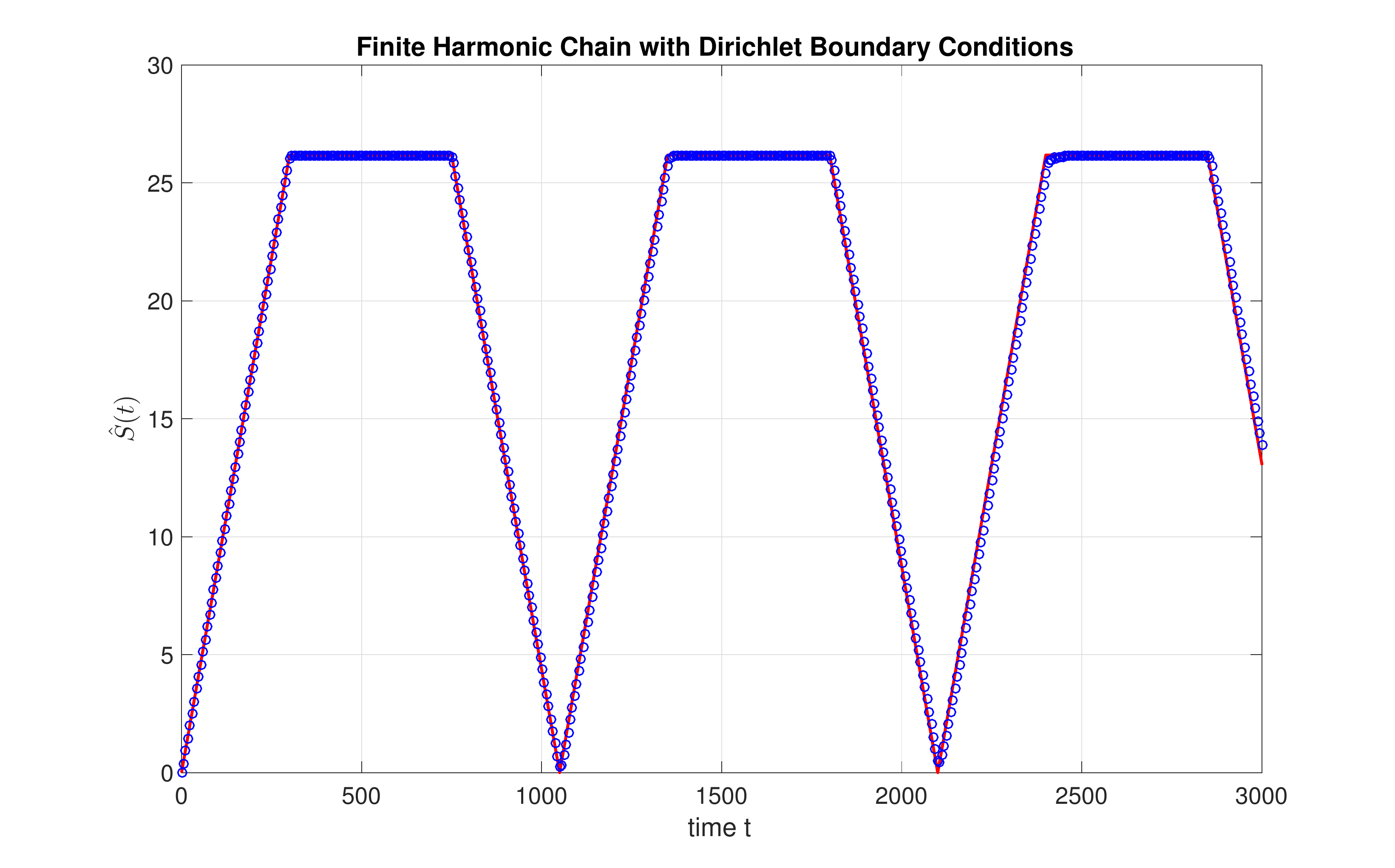}
\caption{\small{Time evolution of the entropy of a finite harmonic chain with Dirichlet boundary conditions on both ends. The region at the end of the chain is of size $2R/a=300$, and the full chain is chosen to be $2L/a=1050$ long in order to avoid any special ratio $L/R$. We follow the time evolution for a long time and find exact revivals. Note that because the region is at the end of the chain, the slope of the curves is half of that in~\eqref{qpPrediction}.}}
\label{fig:finite_harmonic_chain}
\end{center}
\end{figure}

Nevertheless, it is interesting to understand the origin of this $\frac{1}{2} \log t$ behavior, since na\"{i}vely we would have expected saturation in finite time (possibly with $1/t^\#$ power law behavior), and corrections to the quasiparticle model to be suppressed by $\beta/R$. The massless free scalar theory is known to exhibit peculiar logarithmic corrections in the entanglement entropy in static situations due to the presence of a zero mode.\footnote{ It was suggested to us by P. Calabrese that the  behavior we observe here may be related to the $\log\le(\log {R\ov a}\ri)$ correction to the one interval entropy in the vacuum discussed in~\cite{Casini:2005zv,Bianchini:2016mra,Yazdi:2016cxn}.} This motivated us to modify our setup in an attempt at getting rid of the contribution of the soft modes. We observe that  the $\frac{1}{2}\,\log t$ growth disappears if the quench leaves the soft modes in their ground state (i.e. we choose an appropriate $m^2(k)$ in~\eqref{Modified2}),\footnote{This is somewhat subtle, as the zero mode does not have a normalizable ground state.} or if we take the harmonic chain to be finite and consider the interval to be at the end of the chain. Both cases are analyzed in Fig.~\ref{fig:strip_1d_sigmoid_quench} and Fig.~\ref{fig:finite_harmonic_chain}, and the logarithmic growth is clearly gone. 

Based on these results, we can give a heuristic explanation of the $\frac{1}{2} \log t$ behavior based on the dynamics of the zero mode. The following argument was suggested to us by A. Wall. After all the other modes have saturated, we can concentrate on the noncompact zero mode of the scalar field. Its wave function is initially localized, and it spreads under time evolution. The width of the wave function should go as $\sqrt{t}$. Regarding the entropy as the number of available states we immediately obtain the contribution $\log \le(\#\sqrt{t}\ri)=\frac{1}{2}\,\log t+\dots$ to the entropy. That the zero mode contributes the logarithm of its target space volume to the entropy was discussed before in~\cite{Unruh:1990hk,Metlitski:2011pr}. In the smooth mass quench analyzed in Fig.~\ref{fig:strip_1d_sigmoid_quench}, the zero mode is not excited, while for the finite chain analyzed in Fig.~\ref{fig:finite_harmonic_chain}, the zero mode is absent due to the Dirichlet boundary condition, see also~\cite{Yazdi:2016cxn}.

While the detailed exploration of complicated entangling regions and finite volume systems is outside the scope of this work, in Fig.~\ref{fig:finite_harmonic_chain} we follow the time evolution for long times on a finite harmonic chain, where the interval is at one end of the chain.  We have included this geometry to demonstrate that the quasiparticle picture continues to hold in more complicated setups.  The entropy exhibits exact revivals with profile exactly in agreement with the quasiparticle model, which we obtain by mirroring the chain at each end infinite amount of times.  

\subsection{Strips in $2$ and $3$ spatial dimensions}

For the strip geometry in spatial dimensions $d \ge 2$ we can decompose the fields in momentum modes transverse to the entangling surface. Let us denote the entanglement entropy of a massive scalar field of an interval by $S_I(R,t,\beta,m)$, where $\beta$ is the effective temperature in the quench, $2R$ is the width of the interval, and $t$ is the time. Then in $d$ spatial dimensions ($d_\perp=d - 1$ transverse dimensions) the entropy of the strip is (for details see Appendix~\ref{StripAppendix}):
\es{stripEE}{
	S(R,t,\beta) &= A_\perp \int {d^{d-1} k_\perp \ov (2\pi)^{d-1}} \ S_I(R,t,\beta,k_\perp) \,,
} 
where $A_\perp$ is the cross-sectional area of the sides of the strip and $k_\perp$ is the transverse momentum running parallel to the sides of the strip.
We use this formula to compute the entropy numerically. To get a quantity with a well-defined continuum limit we make the subtaction~\eqref{Shat}.

\begin{figure}[!h]
\begin{center}
\includegraphics[scale=0.4]{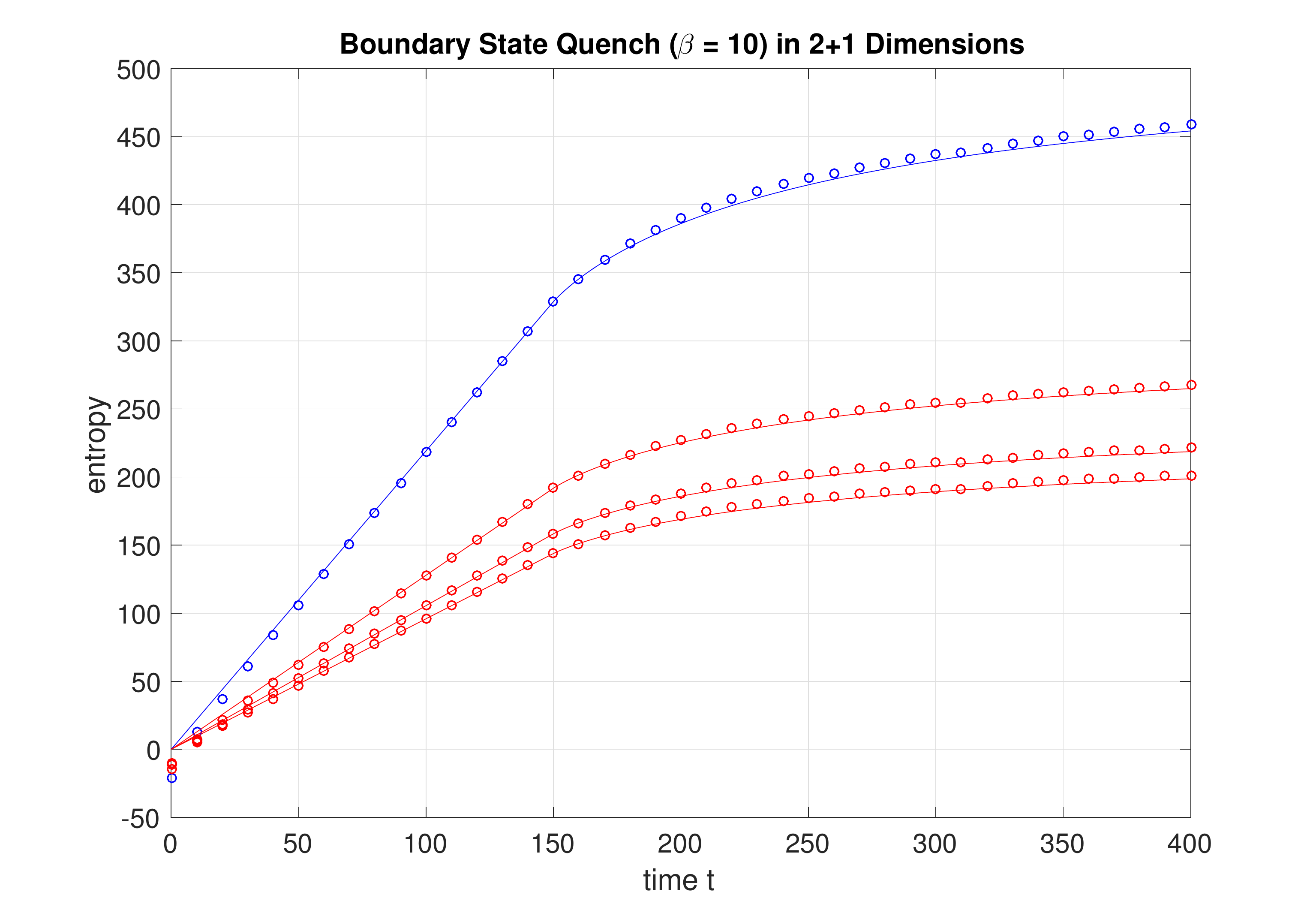}
\includegraphics[scale=0.4]{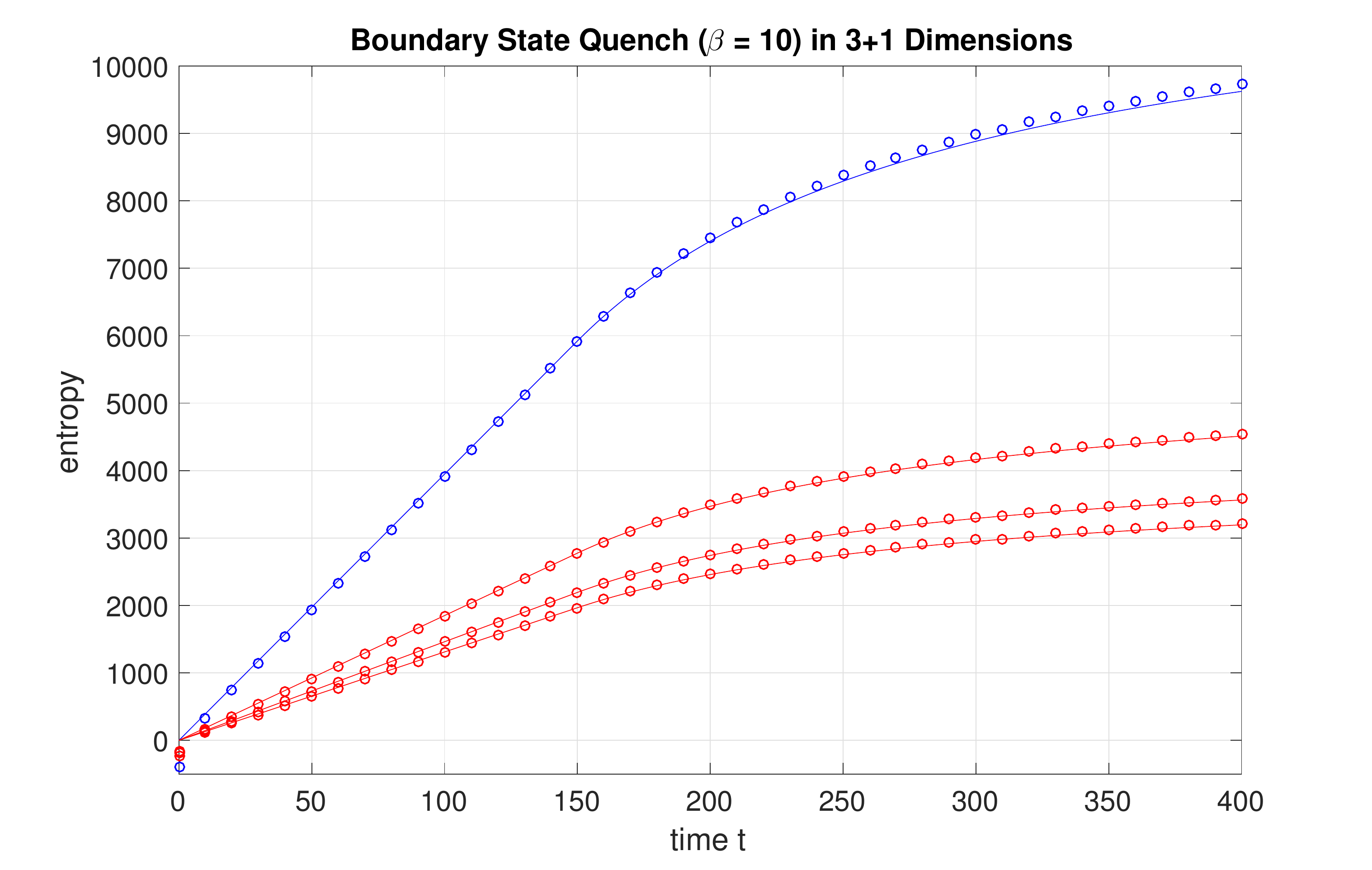}
\caption{\small{Time evolution for a boundary state quench with $\beta = 10 \, a$ of entanglement entropy (blue) and the R\'enyi entropies for $q = 2, 3, 4$ (in red, $q$ increasing from top to bottom), for a strip of width $2R = 300 \, a$.  The numerical data points (circles) are shifted to match the quasiparticle curves (solid lines) at the single point $t = R$. Top figure is for $2+1$ dimensions; bottom figure is for $3+1$ dimensions.}}
\label{fig:strips_2d}
\end{center}
\end{figure}

The results are collected in Fig.~\ref{fig:strips_2d}. In the figures we have also plotted the time evolution of R\'enyi entropies, in addition to the von Neumann entropies.  The results are compared to the predictions of the quasiparticle model~\eqref{qpPrediction}, and precise agreement is found. By precise agreement, we mean up to area law terms subleading in the large region limit, which are not accounted for in the quasiparticle model. In the graphs below we have allowed ourselves to shift the numerical data points by an arbitrary constant to match the quasiparticle prediction. 

We have checked that as we increase the region size this shift scales as the area, and thus is negligible in the limit of large region sizes, see Fig.~\ref{fig:sphere_2d_BSQ_area_law_subtraction} for a demonstration of this in the particular case of a boundary state quench for a spherical geometry, which are discussed in a following section.  We expect similar results for the strip geometries.\footnote{Simulations with truly large region sizes are costly, and do not seem to be necessary to confirm the overall picture.}

The attentive reader may notice some deviation from the quasiparticle curve at early times, $t\sim \beta$ in Fig.~\ref{fig:strips_2d}. Such times do not obey the double scaling limit~\eqref{Scaling3}, hence we do not expect a precise match between the numerical results and quasiparticle curve. In particular until $t\sim \beta$ the entropy grows quadratically~\cite{Calabrese:2005in,Liu:2013iza,Liu:2013qca,2014PhRvB..90t5438N}, while the quasiparticle curve exhibits linear growth~\eqref{LinGrowth}. By smearing the time of origin of the EPR pairs, one can incorporate this quadratic growth into the quasiparticle model~\cite{Casini:2015zua}, but we chose to work with the simplest version of the model, which does not involve any adjustable parameters.\footnote{We do not regard the entropy density $s$ as a fitting parameter of the quasiparticle model, as it can be computed, see Sec.~\ref{sec:Quenches}.}

\begin{figure}[!h]
\begin{center}
\includegraphics[scale=0.4]{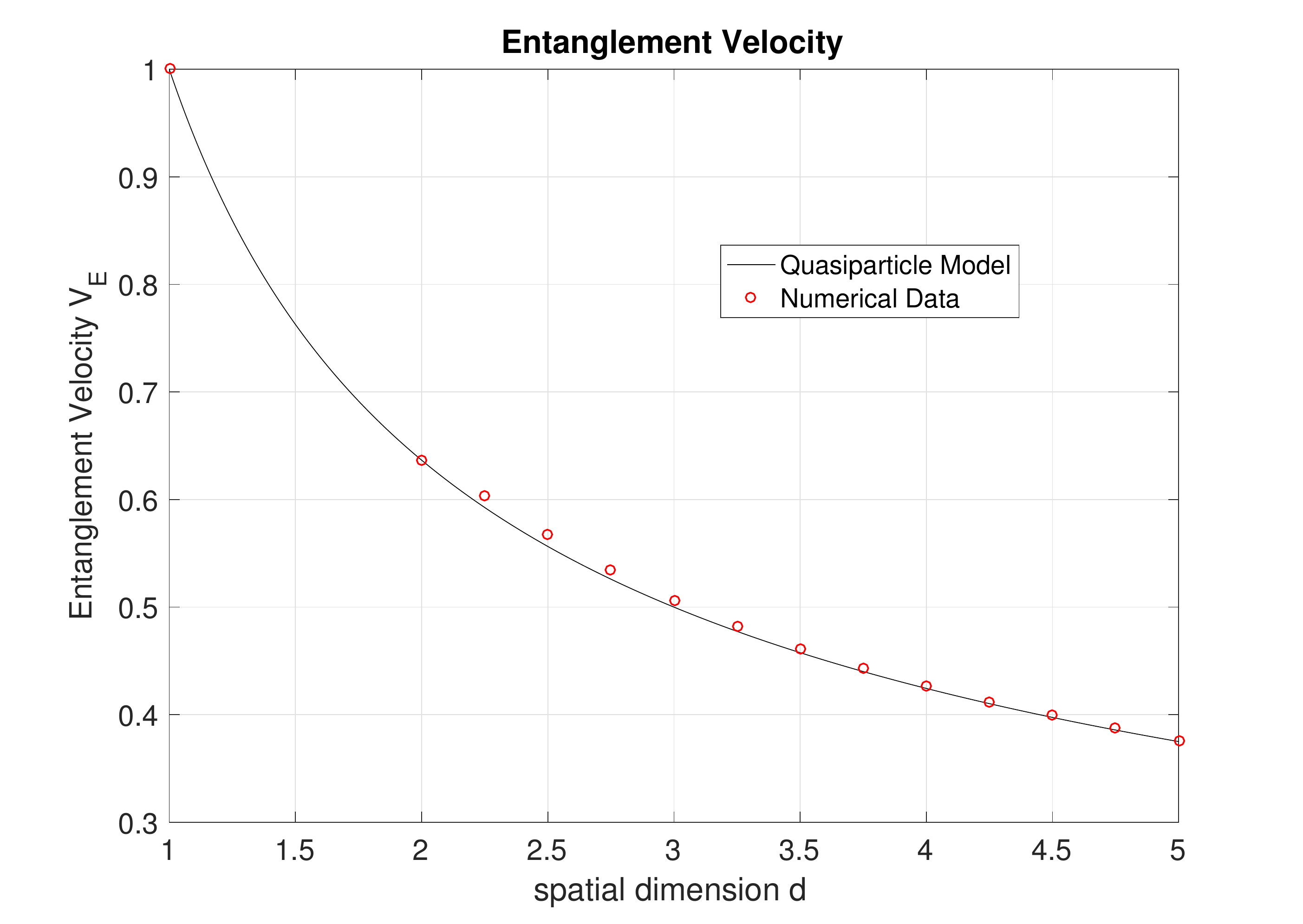}
\caption{\small{The entanglement velocity $v_E$ as extracted from the early time behavior of~\eqref{stripEE} according to the procedure~\eqref{vEstrip}. The numerical results are within $1\%$ of the analytic prediction~\eqref{vEQP}, even for fractional dimensions. } }
\label{fig:strips_entanglement_velocity222}
\end{center}
\end{figure}

Next we focus on an important aspect of the entropy growth, the entanglement velocity defined in~\eqref{LinGrowth}. Because the quasiparticle model predicts exact linear growth until $t=R$, and because at early times we observe more deviation from linearity, we extract $v_E$ from the slope of the curve at $t=R$:
\begin{align}
\label{vEstrip}
	v_E &= {1\ov {2 s \,  A_\perp}} \, {dS(R,t)\ov dt}\Big\vert_{t=R}\,. 
\end{align} 
Numerical results are given in Fig.~\ref{fig:strips_entanglement_velocity222} based on this equation, and they  show very good agreement with the quasiparticle value~\eqref{vEQP} even for fractional dimensions.

\clearpage

\subsection{Spheres in $2$ and $3$ spatial dimensions}

We complete the presentation of the numerical results with the sphere geometry. As discussed in the Introduction, it is important to consider different geometries for the entangling region to confirm the quasiparticle model. The sphere is a particularly nice case to analyze because of its symmetries, which make the numerical computations for large regions possible. See Appendix~\ref{SphereAppendix} for details of the setup.  

\begin{figure}[!h]
\begin{center}
{\includegraphics[scale=0.4]{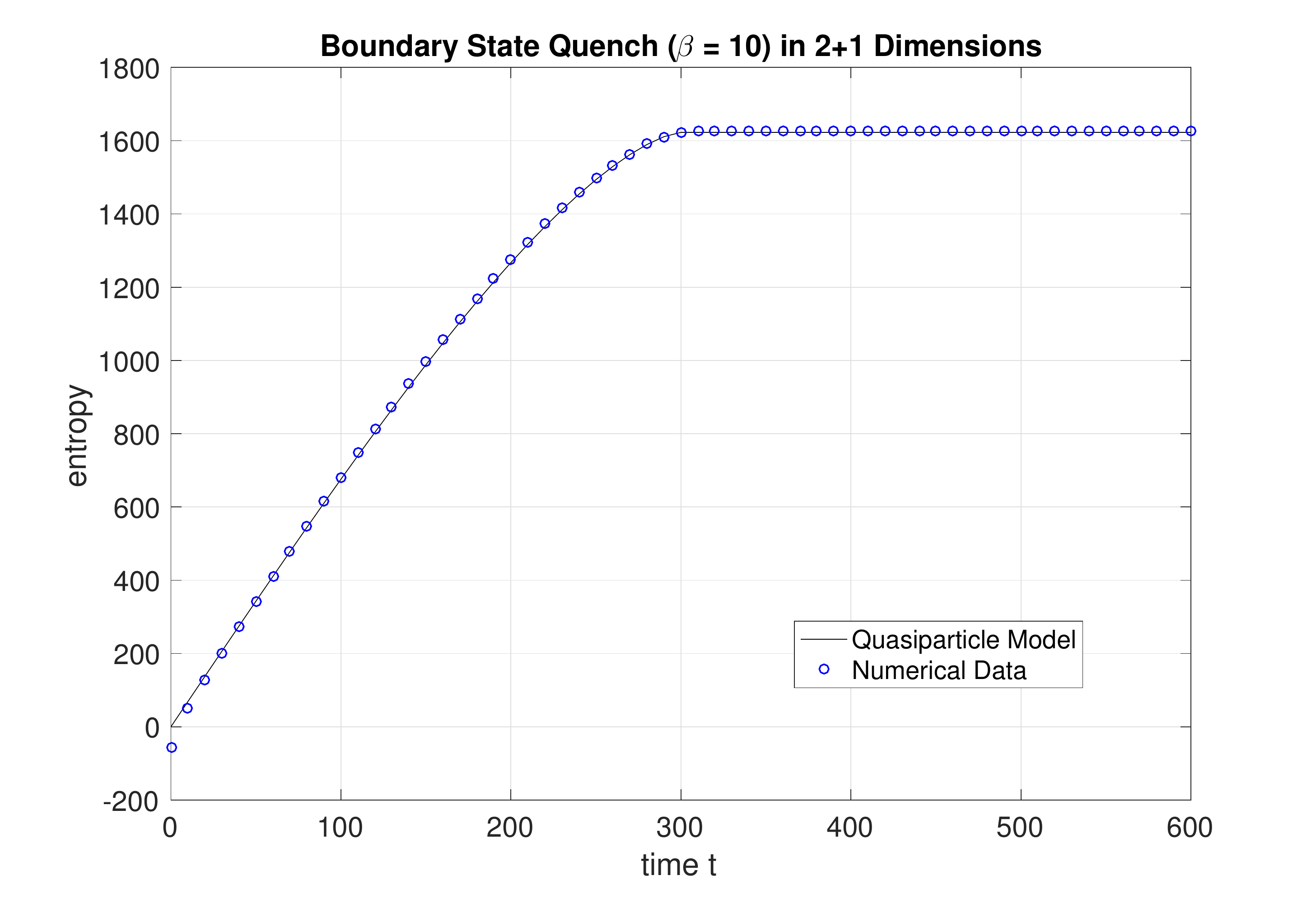}}
{\includegraphics[scale=0.4]{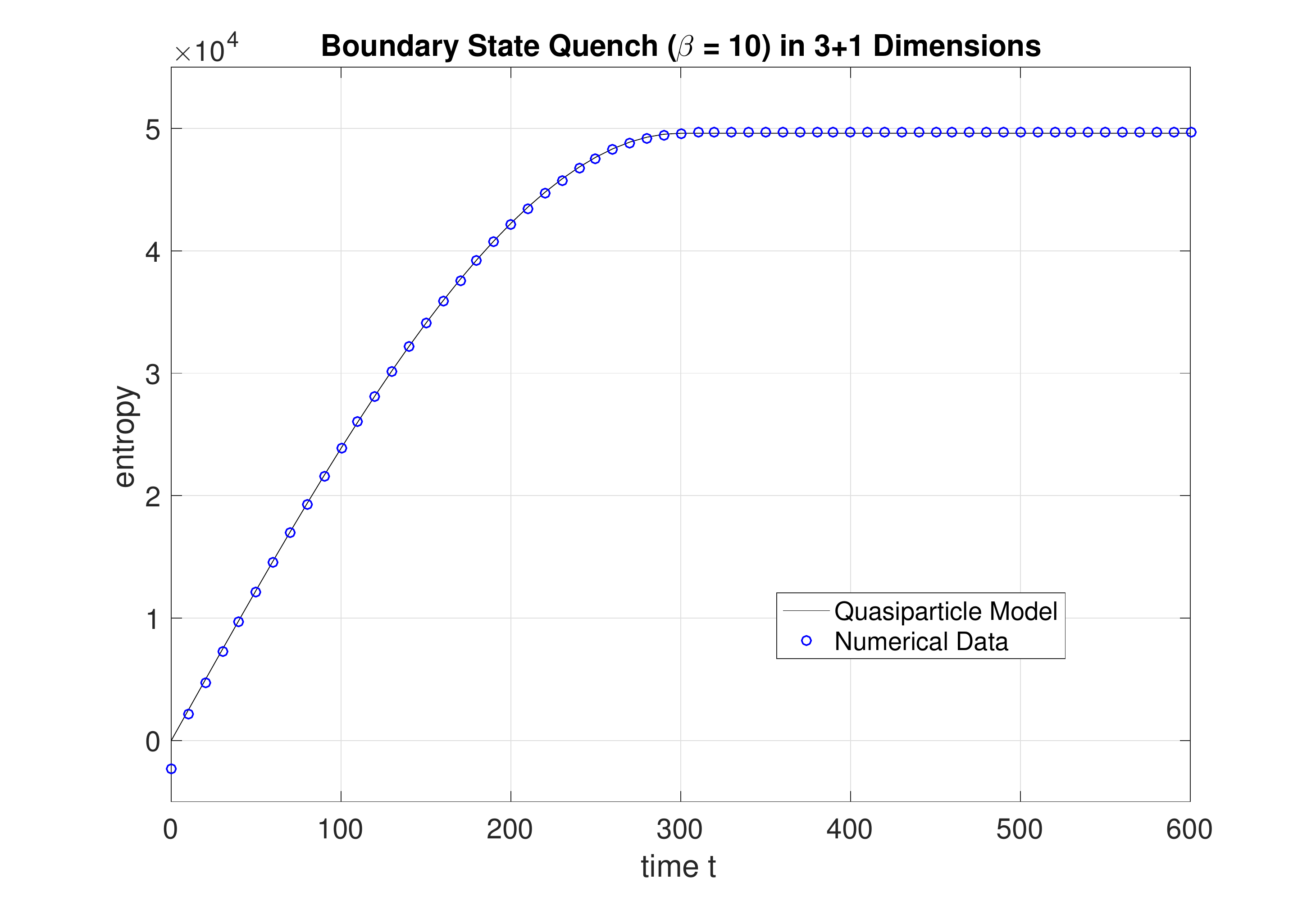}}
\caption{\small{Boundary state quench with $\beta = 10 \, a$ in different dimensions for a sphere of radius $R=300 \, a$ with an arbitrary area law shift to match the quasiparticle curve. The top figure is for $2+1$ dimensions; the bottom is for $3+1$ dimensions.}}
\label{fig:BSQ2dSphere}
\end{center}
\end{figure}

\begin{figure}[!h]
\vspace{-2cm}
\begin{center}
\includegraphics[scale=0.4]{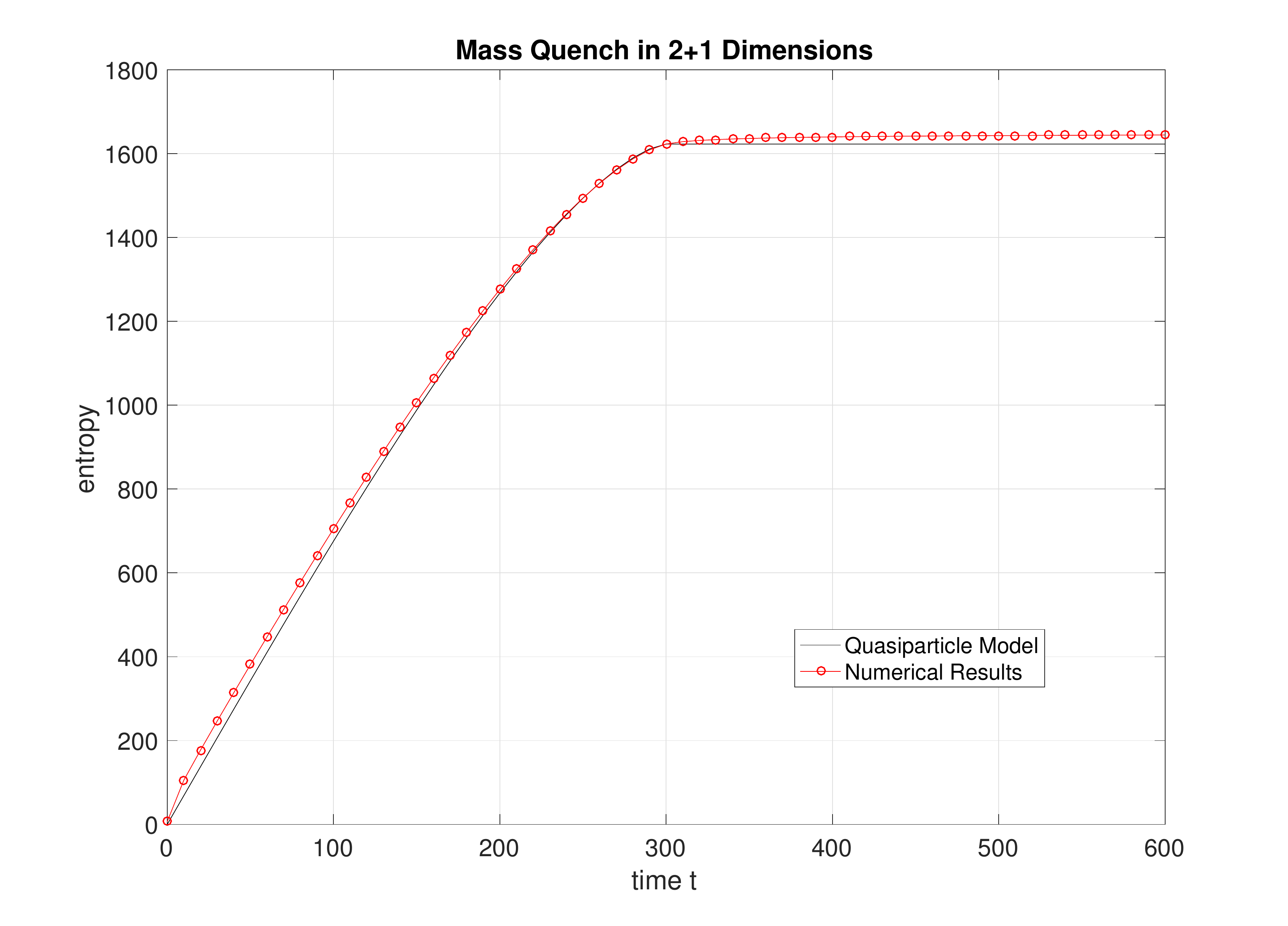}
\vspace{-0.5cm}
\caption{\small{Mass quench in 2+1 dimensions with a sphere of radius $R = 300 \, a$  with an arbitrary area law shift to match the quasiparticle curve.}}
\label{fig:MQ2dSphere}
\end{center}
\end{figure}

In Fig.~\ref{fig:BSQ2dSphere} we plot the results of a boundary state quench in spatial dimensions $d = 2$ and $d = 3$, and in Fig.~\ref{fig:MQ2dSphere} we plot the results of a mass quench in spatial dimensions $d = 2$.  All of these closely match the quasiparticle expectations for all times. Two highlights are that the linear regime is governed by the entanglement velocity~\eqref{vEQP}, and the saturation time is $t_S=R$~\eqref{SatTime}. We note that at late times we again see a logarithmic rise of the entropy after $t_S$, as in the one interval case in $d=1$. This growth is most pronounced on Fig.~\ref{fig:MQ2dSphere}, but the volume law in $d=2$ provides more suppression than in $d = 1$.\footnote{In the $\ell=0$ angular momentum sector we effectively have a massless $d=1$ bosonic chain with a slightly unusual kinetic term, see Appendix~\ref{SphereAppendix}. Thus similar phenomena to the $d=1$ case arising from the zero mode of the scalar are to be expected.} 

\begin{figure}[!h]
\vspace{-0.5cm}
\begin{center}
\includegraphics[scale=0.4]{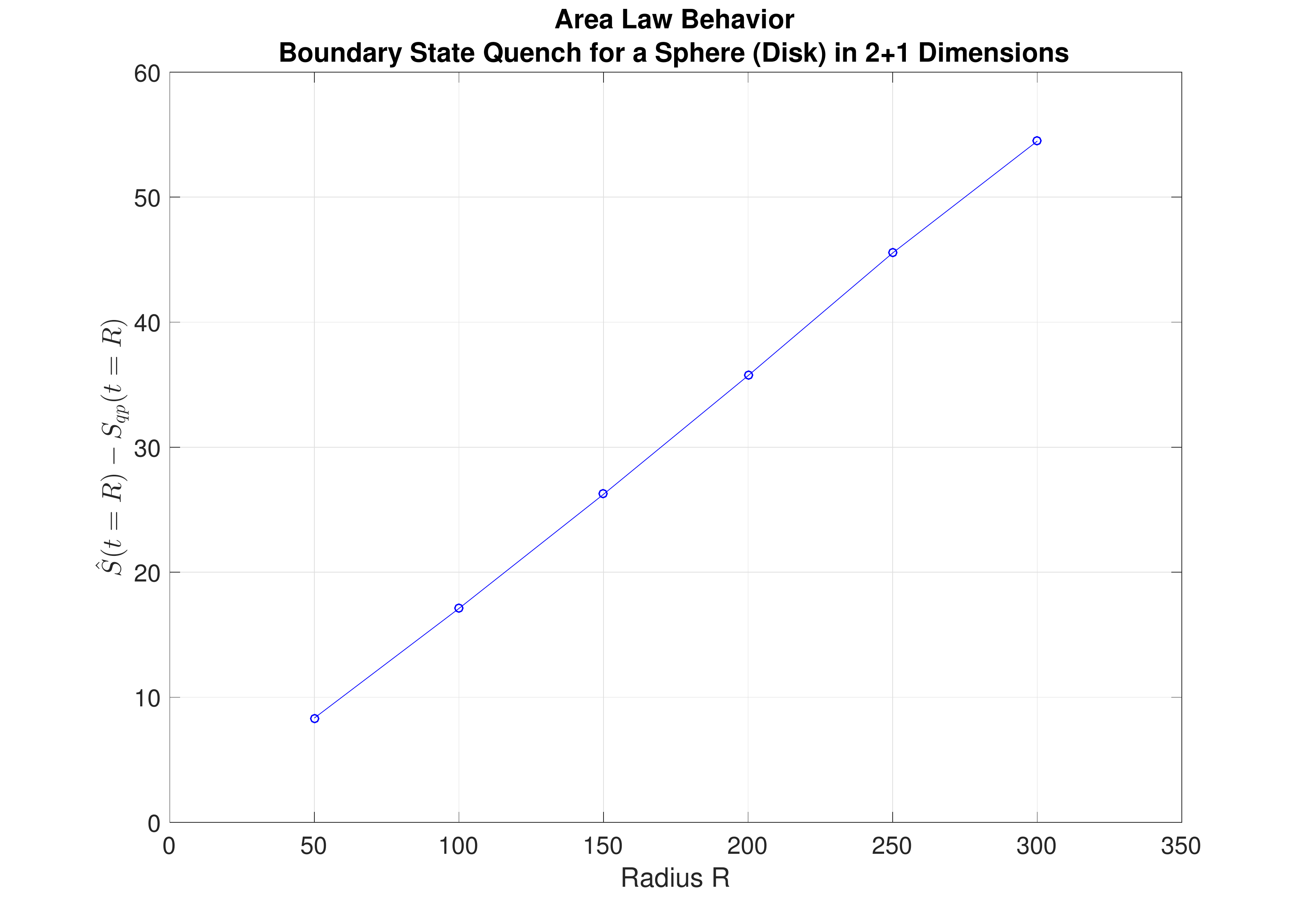}
\vspace{-0.5cm}
\caption{\small{Area law subtraction for a boundary state quench in a spherical (disk) geometry in 2+1 dimensions for different radii $R$ of the disk. Because for spherical geometries $t_S=R$, we require a match between the numerical data points and the quasiparticle curve at $t=R$. Then the shift that we apply to the numerical data is $\hat{S}(t = R) - S_{qp}(t=R)$. It is linear in the radius $R$ of the disk, as expected.}}
\label{fig:sphere_2d_BSQ_area_law_subtraction}
\end{center}
\end{figure}

Finally, for the example of a boundary state quench in $d=2$, in  Fig.~\ref{fig:sphere_2d_BSQ_area_law_subtraction} we demonstrate that the additional shift we apply to the numerical data points to get a closer fit with the quasiparticle model curves obeys the area law, hence it is subleading for large regions.


\clearpage

\section*{Acknowledgments}

We thank P. Calabrese, J. Cardy, H. Casini, N. Lashkari, B. Le Floch, H. Liu, J. Maldacena, and especially A. Wall for useful discussions.
Jordan Cotler is supported by the Fannie and John Hertz Foundation and the Stanford Graduate Fellowship program.  Mark Hertzberg would like to acknowledge support by the Institute of Cosmology at Tufts University.  The research of M\'ark Mezei was supported in part by the U.S. Department of Energy under grant No. DE-SC0016244.    Mark Mueller is supported by the MIT Department of Physics under U.S. Department of Energy grant Contract Number DE-SC00012567.

\appendix

\section{Coordinate systems and mode decompositions } \label{app:apendices}

\subsection{Coordinates for strip geometries}\label{StripAppendix}

Consider a strip geometry in $d+1$ dimensions, where we are interested in tracing over a $d$ dimensional slab of width $2R$ and cross-sectional area $A_\perp$. 

This system factorizes into the physics along the direction $x$ between the sides of the slab and the $d_\perp=d-1$ perpendicular directions we denote by the vector ${\bf x}_\perp$. We write the position vector as
\begin{equation}
	{\bf x} = (x,{\bf x}_\perp) \,.
\end{equation}
In the transverse directions we perform a Fourier transform as follows
\es{}{
    	\tilde\phi(x,{\bf k}_\perp) \equiv \int d^{d-1}x_\perp\,\phi(x,{\bf x}_\perp)\,e^{i{\bf k}_\perp\cdot{\bf x}_\perp} \\
        	\tilde\pi(x,{\bf k}_\perp) \equiv \int d^{d-1}x_\perp\,\pi(x,{\bf x}_\perp)\,e^{i{\bf k}_\perp\cdot{\bf x}_\perp} \,.
}
Inserting this into the Hamiltonian of a massive scalar field (\ref{ContHam}) and using the inverse Fourier theorem, we obtain
\begin{equation}
    	H = {1\over2} \int {d^{d-1}k_\perp\over(2\pi)^{d-1}}\left[|\tilde\pi(x,{\bf k}_\perp)|^2+|\partial_x\tilde\phi(x,{\bf k}_\perp)|^2+(m^2+k_\perp^2)|\tilde\phi(x,{\bf k}_\perp)|^2\right] \,.
\label{Hamp}\end{equation}
The canonical commutation relation for these functions defined as a mix of position space and momentum space is
\begin{equation}
   	[\tilde\phi(x,{\bf k}_\perp),\tilde\pi(x',{\bf k}_\perp')]=i(2\pi)^{d-1}\delta(x-x')\delta^{d-1}({\bf k}_\perp-{\bf k}_\perp')\,.
\end{equation}      
Since this is a free theory, we know that only a single value of momentum ${\bf k}_\perp$ appears in the Hamiltonian in eq.~(\ref{Hamp}). This means we only need the canonical commutation relation when ${\bf k}_\perp'={\bf k}_\perp$. When we take ${\bf k}_\perp'={\bf k}_\perp$ the delta function gives an infrared divergence that is regulated by assuming a finite size transverse region area $A_\perp$, i.e.,
\begin{equation}
   	(2\pi)^{d-1}\delta^{d-1}({\bf k}_\perp-{\bf k}_\perp=0)\to A_\perp\,.
\end{equation}
This motivates re-scaling the fields as follows
\es{}{
        q_{k_\perp}(x)\equiv{1\over\sqrt{A_\perp}}\,\tilde\phi(x,{\bf k}_\perp)\\
        p_{k_\perp}(x)\equiv{1\over\sqrt{A_\perp}}\,\tilde\pi(x,{\bf k}_\perp)\,.
}
The corresponding commutation relation appears canonically normalized for a field dependent only on the $x$ variable, with ${\bf k}_\perp$ just an external parameter
\begin{equation}
      [q_{k_\perp}(x),p_{k_\perp}(x')]=i\,\delta(x-x')\,.
\end{equation} 
The Hamiltonian may then be re-written in terms of these new fields as
\begin{equation}
      	H = A_\perp\int {d^{d-1}k_\perp\over(2\pi)^{d-1}}\,H_1(q_{k_\perp},p_{k_\perp},m^2+k_\perp^2)\,,
\label{HamDecomp}\end{equation}
where $H_1$ is the Hamiltonian of a massive scalar field (of mass $M^2=m^2+k_\perp^2$) in 1+1 dimensions
\begin{equation}
   	H_1(q,p,M^2) = {1\over2}\int dx\left[p^*p+\partial_x q^*\partial_x q+M^2 q^*q\right]\,,
\end{equation}
where in this last equation we have suppressed writing the dependence on the transverse momenta $k_\perp$ to emphasize that this is really just a Hamiltonian defined in 1 spatial dimension with axis running from one side of strip to the other.

The discretization of modes in 1 spatial dimension is rather straightforward, as we now explain. Slightly more complicated discretization that are relevant to the disk or sphere will be discussed in the next subsection. With lattice spacing $a$, the Hamiltonian in 1 spatial dimension is
\begin{equation}
    	H_1(q,p,M^2) = {1\over 2a}\sum_{j=1}^N\left[p(j)^2+\left(q(j)-q(j+1)\right)^2+M_a^2\,q(j)^2\right] \,,
\end{equation}
where the physical lattice points are $x=j\,a$, an IR cutoff is $\IR=N\,a$, and we have defined the dimensionless mass $M_a\equiv M\,a$. We can also impose the conditions $q(N+1)=q(1)$ and $p(N+1)=p(1)$ to impose periodic boundary conditions where necessary. The non-zero elements of the $K$ matrix of eq.~(\ref{HamK}) are  given by
\es{}{
         K^{jj} =& \,2+M_a^2\\
         K^{j,j+1} =& K^{j+1,j}=-1\,,
}
and if periodic boundary conditions are imposed $K^{1N}=K^{N1}=-1$.

Since the different $k_\perp$-modes decouple, and we are only tracing over the strip in the $x$-direction, rather than the ${\bf x}_\perp$ direction, then just as the Hamiltonian decomposes as in (\ref{HamDecomp}) then so too does the entanglement entropy
\begin{equation}
       	S(R,t,m^2)=A_\perp\int {d^{d-1}k_\perp\over(2\pi)^{d-1}}\,S_I(R,t,m^2+k_\perp^2)\,.
\label{SDecomp}\end{equation}
Similar results go through for the boundary state quench, where the entropy depends on the inverse temperature $\beta$, as given in (\ref{stripEE}).

We note that since the entropy $S_I$ is only a function of the magnitude and not the direction of the transverse momenta $k_\perp^2$ in (\ref{SDecomp}), then the angular integration of ${\bf k}_\perp$ is trivial. This gives a factor of the area of the $d-2$ dimensional unit sphere, i.e.,
\begin{equation}
        \int d^{d-1}k_\perp={2\pi^{(d-1)/2}\over\Gamma\left(d-1\over2\right)} \int_0^\infty dk_\perp\,k_\perp^{d-2}\,.
\end{equation}

\subsection{Spherical coordinates and spherical harmonics} \label{SphereAppendix}

In 2+1 dimensions, for a spherical (disk) geometry, we use the Fourier expansions of the field $\phi$ and conjugate momentum $\pi$
\es{}{
	\phi(r,\theta) &= \sum_{\ell = -\infty}^{\infty} \phi_{\ell}(r) \, e^{- i \ell \theta}  \\
	\pi(r,\theta) &= \sum_{\ell = -\infty}^{\infty} \pi_{\ell}(r) \, e^{- i \ell \theta}
}
with Fourier coefficients
\es{}{
	\phi_{\ell}(r) &=  \int_0^{2\pi} \frac{d\theta}{2\pi} \, e^{- i \ell \theta} \phi(r,\theta)  \\
	\pi_{\ell}(r) &=  \int_0^{2\pi} \frac{d\theta}{2\pi} \, e^{- i \ell \theta} \pi(r,\theta) \,.
}
Since $\phi$ is real, $\phi^{*}_{\ell} = \phi_{-\ell}$, and similarly for $\pi$.  Note that the harmonics $\phi_\ell, \pi_\ell$ satisfy the canonical commutation relations
\begin{equation}
	[\phi_{\ell}(r), \pi_{\ell^{\prime}}(r^{\prime})] = \frac{i}{2\pi r} \, \delta_{\ell + \ell^{\prime}} \, \delta(r - r^{\prime} ) \,.
\end{equation}
In terms of these harmonics, the Hamiltonian for a free scalar field of mass $m$ takes the form $H = \sum_{\ell = -\infty}^{\infty} H_{\ell}$, where
\begin{equation}
	H_{\ell} = \int_0^{\infty} 2\pi r dr \  \frac{1}{2} \left[ \pi^{*}_{\ell} \, \pi_{\ell} 
	+ \frac{\partial\phi^{*}_{\ell}}{\partial r}\frac{\partial\phi_{\ell}}{\partial r} 
	+  \left( \frac{\ell^2}{r^2} + m^2 \right) \phi^{*}_{\ell} \, \phi_{\ell}\right]\,.
\end{equation}
If we define the new variables
\es{}{
	q_{\ell}(r) &= \sqrt{2\pi r} \, \phi_{\ell}(r)  \\
	p_{\ell}(r) &= \sqrt{2\pi r} \, \pi_{\ell}(r)  \,,
}
then the canonical commutation relations take the standard form
\begin{equation}
	[q_{\ell}(r), p_{\ell^{\prime}}(r^{\prime})] = i \, \delta_{\ell+\ell^{\prime}} \, \delta(r - r^{\prime} ) \,.
\end{equation}
and the Hamiltonian modes are
\begin{equation}
	H_{\ell} = \int_0^{\infty} dr \  \frac{1}{2} \left[ \, p^{*}_{\ell} \, p_{\ell} 
	+ r \left(\frac{\partial}{\partial r} \frac{q^{*}_{\ell}}{\sqrt{r}} \right) \! \left(\frac{\partial}{\partial r} \frac{q_{\ell}}{\sqrt{r}} \right) 
	+  \left( \frac{\ell^2}{r^2} + m^2 \right) q^{*}_{\ell} \, q_{\ell} \right]\,.
\end{equation}
We may now discretize this Hamiltonian with a uniform lattice in the radial direction:
\begin{equation}
	H_{\ell} = \frac{1}{2a} \sum_{j=1}^N \le[ \, p_{\ell}(j)^2
	+\le(j+\ha\ri) \mkern-4mu \le( \frac{q_{\ell}(j)}{\sqrt{j}} - \frac{q_{\ell}(j+1)}{\sqrt{j+1}} \ri)^{\mkern-4mu 2}
	+ \left( \frac{{\ell}^2}{j^2} + m_a^2 \right) q_{\ell}(j)^2 \ri] \,,
	\label{hscalardisc}
\end{equation}
where $a$ is the lattice spacing and $r=j a$, we introduced an IR cutoff $\IR=Na$, and we have defined the dimensionless mass $m_a\equiv m\,a$. The radius of the disk is taken to be:
\be
	R = \le(n+\ha\ri) a \ . \label{rchoice}
\ee
So in 2+1 dimensions the non-zero elements of the $K$ matrix for the discrete Hamiltonian, which was defined earlier in~\eqref{HamK}, are
\es{defk}{
        K_{\ell}^{11} &= \frac{3}{2} + {\ell}^2+m_a^2  \\
        K_{\ell}^{jj} &= 2 + \frac{{\ell}^2}{j^2} + m_a^2  \\
        K_{\ell}^{j,j+1} &= K_{\ell}^{j+1,j} = -\frac{j+1/2}{\sqrt{j(j+1)}} \,.
}
This matrix and its eigenvalues form the basis for the numerical computations in the correlator method for the entanglement entropy.  In the counting for the different modes in the entropy calculations, we must sum over all $\ell \ge 0$, with the $\ell=0$ mode getting a factor of  1 and the other modes a factor of 2:
\es{Ssum2d}{
S=S_0+\sum_{\ell=1}^\infty 2\,S_\ell\,.
}

In 3+1 dimensions, for a spherical geometry, the development is similar to 2+1 dimensions, with some important differences.  We use the expansions of the field $\phi$ and conjugate momentum $\pi$ in terms of spherical harmonics
\es{}{
	\phi(r,\Omega) &= \sum_{\ell,m} \phi_{\ell m}(r) \, Y_{\ell m}(\Omega)  \\
	\pi(r,\Omega) &= \sum_{\ell,m} \pi_{\ell m}(r) \, Y_{\ell m}(\Omega)
}
with inversion formulas
\es{}{
	\phi_{\ell}(r) &=  \int \frac{d\Omega}{4\pi} \, Y^{*}_{\ell m}(\Omega) \, \phi(r,\Omega)  \\
	\pi_{\ell}(r) &=  \int \frac{d\Omega}{4\pi} \, Y^{*}_{\ell m}(\Omega) \, \pi(r,\Omega)\,.
}
In terms of these harmonics, the Hamiltonian for a free scalar field of mass $m$ takes the form $H = \sum_{\ell m} H_{\ell m}$, where
\begin{equation}
	H_{\ell m} = \int_0^{\infty} 4\pi r^2 dr \  \frac{1}{2} \left[ \pi^{*}_{\ell m} \pi_{\ell m} 
	+ \frac{\partial\phi^{*}_{\ell m}}{\partial r}\frac{\partial\phi_{\ell m}}{\partial r} 
	+  \left( \frac{\ell (\ell + 1)}{r^2} + m^2 \right) \phi^{*}_{\ell m} \, \phi_{\ell m}\right]\,.
\end{equation}
If we define the new variables
\es{}{
	q_{\ell m}(r) &= \sqrt{4\pi} \, r \, \phi_{\ell m}(r)  \\
	p_{\ell m}(r) &= \sqrt{4\pi} \, r \, \pi_{\ell m}(r)  \,,
}
the Hamiltonian modes are
\begin{equation}
	H_{\ell m} = \int_0^{\infty} dr \  \frac{1}{2} \left[ \, p^{*}_{\ell m} \, p_{\ell m} 
	+ r \left(\frac{\partial}{\partial r} \frac{q^{*}_{\ell m}}{r} \right) \mkern-4mu \left(\frac{\partial}{\partial r} \frac{q_{\ell m}}{r} \right)
	+  \left( \frac{\ell (\ell + 1)}{r^2} + m^2 \right) q^{*}_{\ell m} \, q_{\ell m} \right] \,.
\end{equation}
Discretizing this Hamiltonian with a uniform lattice in the radial direction:
\begin{equation}
	H_l = \frac{1}{2a} \sum_{j=1}^N \mkern-2mu \le[ \, p_{\ell m}(j)^2  \mkern-2mu
	+ \mkern-4mu \le( j+\ha \ri)^{\! 2} \mkern-8mu \le( \frac{q_{\ell m}(j)}{j} - \frac{q_{\ell m}(j+1)}{j+1} \ri)^{\mkern-8mu 2} \mkern-4mu
	+ \mkern-4mu \left( \frac{\ell (\ell + 1)}{j^2} + m_a^2 \right) \mkern-4mu q_{\ell m} (j)^2 \ri]\,,
	\label{hscalardisc2}
\end{equation}
where again $a$ is the lattice spacing, $r=j a$ and the radius of the sphere, and the radius of the sphere is given as in (\ref{rchoice}).  So, in 3+1 dimensions the non-zero elements of the $K$ matrix for the discrete Hamiltonian are
\es{defk3}{
        K_{\ell}^{11} &= \frac{9}{4} + \ell (\ell + 1) + m_a^2  \\
        K_{\ell}^{jj} &= 2 + \frac{1}{2 j^2} + \frac{\ell (\ell + 1)}{j^2} + m_a^2  \\
        K_{\ell}^{j,j+1} &= K_{\ell}^{j+1,j} = -\frac{(j+1/2)^2}{j(j+1)}\,.
}
As in 2+1 dimensions, this matrix and its eigenvalues form the basis for the numerical computations in the correlator method for the entanglement entropy. The entropy is finally given by the sum over the entropies coming from each angular momentum sector:
\es{Ssum3d}{
S=\sum_{\ell=0}^\infty (2\ell+1)\,S_\ell\,.
}

\bibliographystyle{ssg}
\bibliography{EE_free_scalar}

\end{document}